\documentclass[fleqn,usenatbib,letterpaper]{mnras_arXiv}

\usepackage{newtxtext,newtxmath}

\usepackage[T1]{fontenc}
\usepackage{ae,aecompl}
\usepackage{threeparttable}

\usepackage{graphicx}	
\usepackage{amsmath}	
\usepackage{amssymb}	
\usepackage{xcolor}

\title[The ram pressure stripped radio tails of galaxies]{The ram pressure stripped radio tails of galaxies in the Coma cluster}

\author[H. Chen et al.]{
Hao Chen,$^{1,2}$\thanks{E-mail: HaoChen.cluster@gmail.com}
\ Ming Sun,$^{1}$\thanks{E-mail: ms0071@uah.edu}
\ Masafumi Yagi,$^{3,4}$%
\ Hector Bravo-Alfaro,$^{5}$%
\ Elias Brinks,$^{6}$%
\newauthor 
\ Jeffrey Kenney,$^{7}$%
\ Francoise Combes,$^{8}$%
\ Suresh Sivanandam,$^{9,10}$%
\ Pavel Jachym,$^{11}$%
\newauthor 
\ Matteo Fossati,$^{12,13}$%
\ Giuseppe Gavazzi,$^{14}$%
\ Alessandro Boselli,$^{15}$%
\ Paul Nulsen,$^{16,17}$%
\newauthor 
\ Craig Sarazin,$^{18}$%
\ Chong Ge,$^{1}$%
\ Michitoshi Yoshida,$^{19,20}$%
\ and Elke Roediger,$^{21}$%
\\
$^{1}$Department of Physics \& Astronomy, University of Alabama in Huntsville, 301 Sparkman Drive, Huntsville, AL 35899, USA\\
$^{2}$Department of Astronomy, University of Cape Town, Private Bag X3, Rondebosch 7701, South Africa\\
$^{3}$Optical and Infrared Astronomy Division, National Astronomical Observatory of Japan, Mitaka, Tokyo, 181-8588, Japan\\
$^{4}$Graduate School of Science and Engineering, Hosei University, 3-7-2, Kajinocho, Koganei, Tokyo, 184-8584, Japan\\
$^{5}$Departamento de Astronom\'\i a, Universidad de Guanajuato, Apdo.\ Postal 144, Guanajuato 36000, Mexico\\
$^{6}$Centre for Astrophysics Research, University of Hertfordshire, College Lane, Hatfield AL10 9AB, UK\\
$^{7}$Yale University Astronomy Department, P.O. Box 208101, New Haven, CT 06520-8101, USA\\
$^{8}$Observatoire de Paris, LERMA, College de France, CNRS, PSL Univ, Sorbonne University, UPMC F-75014 Paris, France\\
$^{9}$Department of Astronomy and Astrophysics, University of Toronto, 50 St. George St, Toronto, ON, Canada\\
$^{10}$Dunlap Institute of Astronomy and Astrophysics, University of Toronto, 50 St. George St, Toronto, ON, Canada\\
$^{11}$Astronomical Institute, Academy of Sciences of the Czech Republic, Bo\v cn\' \i \ II 1401, 141 31 Prague 4, Czech Republic\\
$^{12}$Dipartimento di Fisica G. Occhialini, Universit{\"a} degli Studi di Milano Bicocca, Piazza della Scienza 3, I-20126 Milano, Italy\\
$^{13}$Institute for Computational Cosmology and Center for Extragalactic Astronomy, Durham University, South Road, Durham DH1 3LE, UK\\
$^{14}$Universit\`a degli Studi di Milano-Bicocca, Piazza della Scienza 3, 20126 Milano, Italy\\
$^{15}$Aix-Marseille Univ., CNRS, CNES, LAM, Marseille, France\\
$^{16}$Harvard-Smithsonian Center for Astrophysics, 60 Garden Street, Cambridge, MA 02138, USA\\
$^{17}$ICRAR, University of Western Australia, 35 Stirling Hwy, Crawley, WA 6009, Australia\\
$^{18}$Department of Astronomy, University of Virginia, Charlottesville, VA 22904, USA\\
$^{19}$Subaru Telescope, National Astronomical Observatory of Japan, 650 North A'ohoku Place, Hilo, HI 96720 USA\\
$^{20}$Hiroshima Astrophysical Science Center, Hiroshima University, Higashi-Hiroshima, Hiroshima 739-8526, Japan\\
$^{21}$Milne Centre for Astrophysics, Department of Physics $\&$ Mathematics, University of Hull, Hull, HU6 7RX, UK\\
}

\date{Accepted XXX. Received YYY; in original form ZZZ}

\pubyear{2020}

\begin{document}
\label{firstpage}
\pagerange{\pageref{firstpage}--\pageref{lastpage}}
\maketitle

\begin{abstract}
Previous studies have revealed a population of galaxies in galaxy clusters with ram pressure stripped (RPS) tails of gas and embedded young stars. We observed 1.4\,GHz continuum and HI emission with the Very Large Array in its B--configuration in two fields of the Coma cluster to study the radio properties of RPS galaxies. The best continuum sensitivities in the two fields are 6 and 8 $\mu$Jy per 4$\arcsec$ beam respectively, which are 4 and 3 times deeper than those previously published.
Radio continuum tails are found in 10 (8 are new) out of 20 RPS galaxies, unambiguously revealing the presence of relativistic electrons and magnetic fields in the stripped tails.  
Our results also hint that the tail has a steeper spectrum than the galaxy. 
The 1.4\,GHz continuum in the tails is enhanced relative to their H$\alpha$ emission by a factor of $\sim$ 7 compared to the main bodies of the RPS galaxies.
The 1.4\,GHz continuum of the RPS galaxies is also enhanced relative to their IR emission by a factor of $\sim$ 2 compared to star-forming galaxies.
The enhancement is likely related to ram pressure and turbulence in the tail. 
We furthermore present HI detections in three RPS galaxies and upper limits for the other RPS galaxies. 
The cold gas in D100's stripped tail is dominated by molecular gas, which is likely a consequence of the high ambient pressure. 
No evidence of radio emission associated with ultra--diffuse galaxies is found in our data.
\end{abstract}



 \section{Introduction}

\begin{figure*}
    \includegraphics[width=2.1\columnwidth]{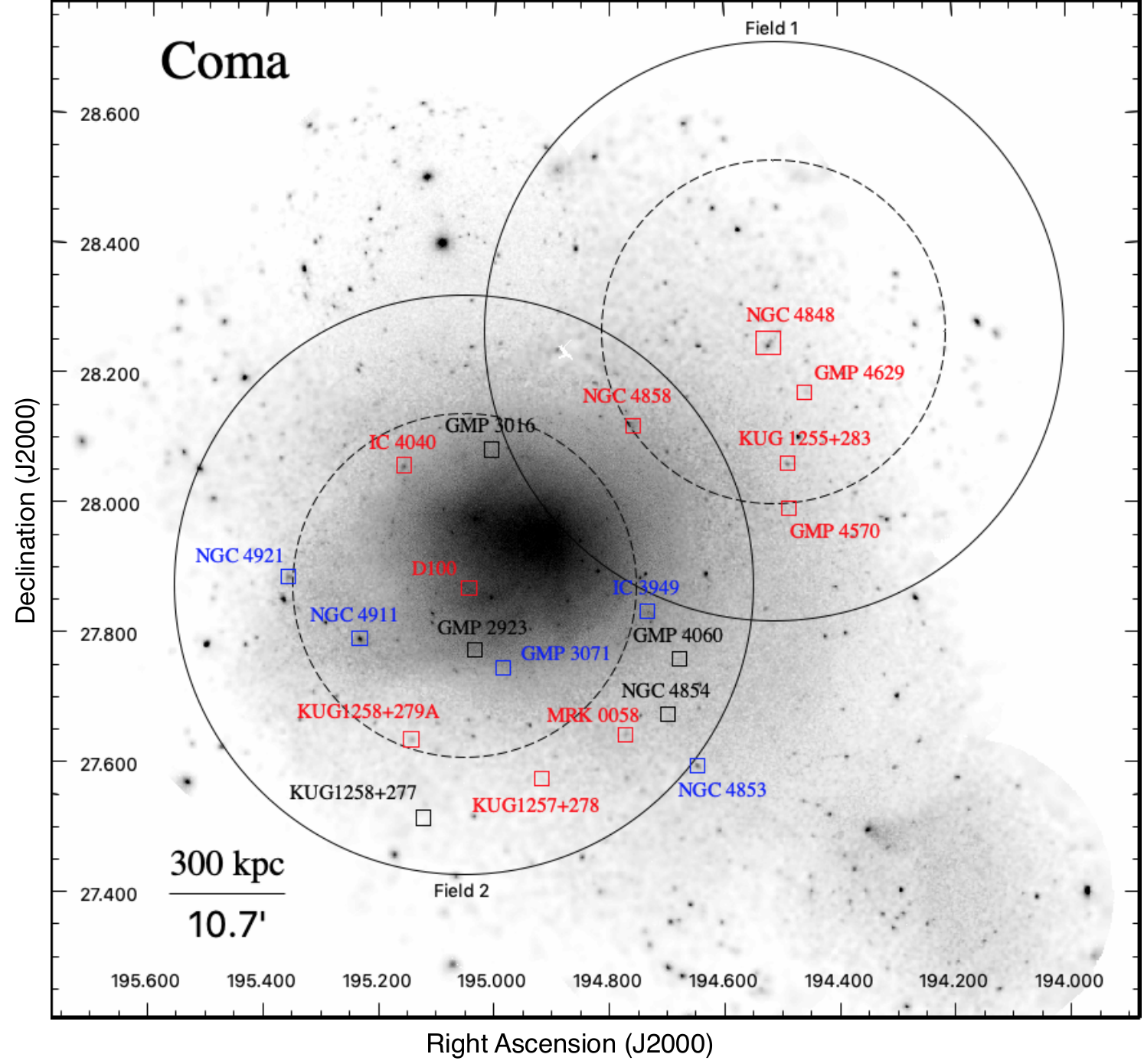}
    \vspace{-0.5cm} 
    \caption{{\em XMM-Newton} 0.4 - 1.3 keV image of the Coma cluster \citep{Snowden2008A&A...478..615S}, with positions of the RPS galaxies marked \citep{Smith2010MNRAS.408.1417S,Yagi2010AJ....140.1814Y,Kenney2015AJ....150...59K}. 
    The RPS galaxies with radio continuum detection in both the galaxy and tail (red), only in the galaxy (blue), and without detection (black) are marked with boxes. 
    The solid and dashed circles show the 10\% and 50\% response (26.77$\arcmin$ and 15.88$\arcmin$ in radius) of the two  VLA fields. 
    }
    \label{figure:coma}
\end{figure*}

Ram pressure stripped (RPS) galaxies are characterized by gas being stripped from the affected galaxy by the intracluster medium \citep[ICM, e.g.,][]{Gunn1972ApJ...176....1G,Nulsen1982MNRAS.198.1007N}.
Star formation (SF) can be triggered by ram pressure at the early interaction stage by compression of interstellar medium (ISM), as shown in observations and simulations \citep[e.g.,][]{Koopmann2004ApJ...613..866K,Crowl2006AAS...20921111C}.
Then, as the cold ISM is depleted, the galactic SF will be quenched \citep[e.g.,][]{Quilis2000Sci...288.1617Q,Boselli2016A&A...596A..11B}. Thus, ram pressure stripping is an important process affecting galaxy evolution in rich environments like galaxy groups and clusters.
The evolution of the stripped ISM is a significant area of research.
The mixing of the stripped cold ISM with the hot ICM will produce a multi-phase gas \citep[e.g.,][]{Sun2007ApJ...671..190S,Ferland2009MNRAS.392.1475F,Jachym2019ApJ...883..145J}. 
Some of the stripped ISM can turn into stars in the galactic halo and the intracluster space \citep[e.g.,][]{Cortese2007MNRAS.376..157C,Sun2007ApJ...671..190S,Owers2012ApJ...750L..23O,Ebeling2014ApJ...781L..40E,Cramer2019ApJ...870...63C}, especially in the high ICM-pressure environment \citep[e.g.,][]{Sun2010ApJ...708..946S}.
Thus, stripped tails emerge as ideal targets to study this multi-phase medium and SF conditions in an extreme environment.

Ram pressure stripped tails are observed
in X-rays \citep[e.g.,][]{Sun2006ApJ...637L..81S,Sun2010ApJ...708..946S,Zhang2013ApJ...777..122Z}, far-ultraviolet \citep[FUV, e.g.][]{Boissier2012A&A...545A.142B}, H$\alpha$ \citep[e.g.,][]{Gavazzi2001ApJ...563L..23G,Gavazzi2017A&A...606A.131G,Gavazzi2018A&A...618A.130G,Sun2007ApJ...671..190S,Yagi2007ApJ...660.1209Y,Yagi2017ApJ...839...65Y,Yoshida2008ApJ...688..918Y,Yoshida2012ApJ...749...43Y,Smith2010MNRAS.408.1417S,Yagi2010AJ....140.1814Y,Yagi2013ApJ...778...91Y,Fossati2012A&A...544A.128F,Fossati2016MNRAS.455.2028F,Fossati2018A&A...614A..57F,Fumagalli2014MNRAS.445.4335F,Boselli2016A&A...587A..68B,Boselli2018A&A...615A.114B,Bellhouse2017ApJ...844...49B}, warm H$_{2}$ \citep[e.g.,][]{Sivanandam2010ApJ...717..147S,Sivanandam2014ApJ...796...89S}, CO \citep[e.g.,][]{Jachym2013A&A...556A..99J,Jachym2014ApJ...792...11J,Jachym2017ApJ...839..114J,Scott2013MNRAS.429..221S,Scott2015MNRAS.453..328S,Verdugo2015A&A...582A...6V,Moretti2018MNRAS.480.2508M} and HI \citep[e.g.,][]{Kenney2004AJ....127.3361K,Kenney2014ApJ...780..119K,Oosterloo2005A&A...437L..19O,Chung2007ApJ...659L.115C,Chung2009AJ....138.1741C,Scott2010MNRAS.403.1175S,Scott2012MNRAS.419L..19S,Abramson2014AJ....147...63A,Ramatsoku2019MNRAS.487.4580R,Serra2019A&A...628A.122S,Deb2020arXiv200404754D}.
Extensive simulations
\citep[e.g.,][]{Quilis2000Sci...288.1617Q,Roediger2008MNRAS.388..465R,Ruszkowski2012atp..prop...17R} show that stripping has a significant impact on galaxy evolution (e.g., disk truncation, the formation of flocculent arms,
the transformation of dwarf galaxies).
SF in the stripped tail has also been seen in simulations \citep[e.g.,][]{Kapferer2009A&A...499...87K,Tonnesen2010ApJ...709.1203T,Tonnesen2012MNRAS.422.1609T,Roediger2014MNRAS.443L.114R}.

\begin{table*}
	\caption{{VLA} 1.4 GHz observations in  B-configuration}
	\begin{tabular}{cccccccc} 
		\hline
		Field & R.A. (J2000) & Decl. (J2000) &Obs. date &Total time (on-source) & Continuum beam & HI beam & Central source\\
		\hline
		1	& 12h58m02.65s & +28$^\circ$15\arcmin50.0\arcsec &  2016, Jun 1--11 & 5 $\times$ 3 h (12.1 h) & $4.1\arcsec\times 3.5\arcsec$ &$6.2\arcsec\times 5.2\arcsec$ & NGC~4848\\
        2	& 13h00m12.46s & +27$^\circ$52\arcmin23.4\arcsec & 2016, Jun 5--10 & 2 $\times$ 4 h (6.3 h) & $4.7\arcsec\times 3.7\arcsec$ &$7.5\arcsec\times 5.2\arcsec$ & D100\\
		\hline
	\end{tabular}
    \label{tab:observation_table}
\end{table*}

\begin{table*}
	\caption{Coma RPS galaxies studied in this paper}
	\begin{tabular}{ccccccccc} 
		\hline
		NO. & Galaxy & R.A. (J2000) & Decl. (J2000) & Velocity & rms  & 1.4 GHz Flux Density & Field\\
            &             & [h m s] & [$^\circ$\ $\arcmin$\ $\arcsec$]                    & $[{\rm km\ s^{-1}}]$ & $[{\rm \mu Jy\ beam^{-1}}]$&   $[{\rm mJy}]$  & \\
		\hline
	1	&NGC~4848 (GMP~4471)    & 12 58 05.59  &  28 14 33.6  & 7184 & 5.94 & $23.85\pm0.11$ & 1\\
	2  &KUG 1255+283 (GMP~4555) & 12 57 57.74 &  28 03 42.1  & 8136 & 8.04 & $6.76\pm0.08$  & 1\\
	3	&NGC 4858 (GMP~3816)    & 12 59 02.11 &  28 06 56.4  & 9416 & 9.85 & $8.73\pm0.13$  & 1\\
    4	&GMP 4570 	 & 12 57 56.81  &  27 59 30.6  & 4565 & 9.92 & $0.55\pm0.06$  & 1\\
    5	&GMP 4629    & 12 57 50.27 &  28 10 13.7  & 6918 & 5.73 & $0.03\pm0.01$  & 1\\
    6 &D100 (MRK~0060, GMP~2910)    & 13 00 09.15  &  27 51 59.4  & 5316 & 8.05 & $1.12\pm0.03$  & 2\\
    7 &KUG 1258+279A (GMP~2599) & 13 00 33.70  &  27 38 15.6  & 7485 & 13.8 & $3.70\pm0.09$  & 2\\
    8	&MRK 0058 (GMP~3779)    & 12 59 05.30  &  27 38 39.6  & 5419 & 19.5 & $3.29\pm0.15$  & 2\\
    9	&IC 4040 (GMP~2559)      & 13 00 37.91  &  28 03 28.0  & 7675 & 10.7 & $16.31\pm0.10$ & 2\\
    10&KUG 1257+278 (GMP~3271)  & 12 59 39.82  &  27 34 35.9  & 5011 & 17.7 & $0.51\pm0.07$  & 2\\
    11	&IC 3949$^*$ (GMP~3896)      & 12 58 55.89  &  27 49 59.9  & 7526 & 14.5 & $2.37\pm0.09$  & 2\\
    12	&GMP 3071$^*$     & 12 59 56.15  &  27 44 47.3  & 8920 & 9.04 & $0.05\pm0.02$  & 2\\
    13	&NGC 4853$^*$ (GMP~4156)     & 12 58 35.20  &  27 35 47.1  & 7688 & 40.8 & $1.35\pm0.15$  & 2\\
    14	&NGC 4911$^*$ (GMP~2374)     & 13 00 56.08  &  27 47 26.9  & 7985 & 10.4 & $15.08\pm0.23$ & 2\\ 
    15	&NGC 4921$^*$ (GMP~2059)     & 13 01 26.15  &  27 53 09.5  & 5470 & 14.6 & $0.44\pm0.05$  & 2\\
    16	&GMP 4060$^*$    & 12 58 42.60  &  27 45 38.0  & 8686 & 20.8 & $<0.083$            & 2\\
    17	&NGC 4854$^*$ (GMP~4017)     & 12 58 47.44  &  27 40 29.3  & 8383 & 24.6 & $<$ 0.098            & 2\\
    18	&GMP3016$^*$      & 13 00 01.08  &  28 04 56.2  & 7765 & 10.8 & $<$ 0.043            & 2\\
    19	&GMP 2923$^*$     & 13 00 08.07  &  27 46 24.0  & 8672 & 8.54 & $<$ 0.034            & 2\\
    20&KUG 1258+277$^*$ (GMP~2640)   & 13 00 29.23  &  27 30 53.7  & 7395 & 23.6 & $<$ 0.094            & 2\\
        \hline
	\end{tabular}
    \begin{tablenotes}
    \item {\sl Note:}
    The properties of 20 RPS galaxies from \citet{Smith2010MNRAS.408.1417S}, \citet{Yagi2010AJ....140.1814Y} and \citet{Kenney2015AJ....150...59K} are listed. R.A., Decl., and optical velocity are from LEDA \citep{Makarov2014A&A...570A..13M}, except for GMP3016 whose velocity is from NED as there is no LEDA value. The local rms and continuum flux density (or 4$\sigma$ upper limit) measured by PyBDSF are also shown for each galaxy. The rms varies within the same field because of the primary beam correction. The synthesised beam sizes are shown in Table~\ref{tab:observation_table}. Galaxies with an asterisk have no RPS tails detected in radio continuum.
    \end{tablenotes}
    \label{tab:tail_sources_table}
\end{table*}

\begin{table*}
	\caption{Radio continuum flux densities of RPS galaxies and their tails}
	\begin{tabular}{cccccccccc} 
		\hline
		NO. & Region identification & Component & 1.4 GHz Flux Density & Spectral Index & Sky Area & Tail Length \\
                 &  &      & [mJy] & & [arcsec$^2$] & [kpc]   \\
		\hline
		1	&NGC~4848a      & nucleus    & $9.495\pm0.011$   & $-0.55\pm0.01$    & 58.2 &  -\\
		1	&NGC~4848b      & galaxy    & 13.207 $\pm$ 0.041  & $-$0.82 $\pm$ 0.02    & 769.3 &  -\\
        1   &NGC~4848c      & tail      & 0.851 $\pm$ 0.032   & $-$0.97 $\pm$ 0.33    & 472.3 & 12.6\\
        2   &KUG1255+283a   & galaxy	& 5.476 $\pm$ 0.032   & $-$1.03 $\pm$ 0.05    & 256.0 &  -\\
        2   &KUG1255+283b   & tail      & 1.366 $\pm$ 0.038   & $-$1.36 $\pm$ 0.25    & 368.6 &  9.1 \\
        3	&NGC4858a       & galaxy    & 6.865 $\pm$ 0.035   & $-$1.32 $\pm$ 0.05    & 209.3 &  -\\
        3   &NGC4858b       & tail      & 1.245 $\pm$ 0.045   & $-$2.25 $\pm$ 0.49    & 341.8 &  10.8 \\
        4	&GMP 4570a      & galaxy    & 0.297 $\pm$ 0.023   & $-$1.08 $\pm$ 0.82    &  84.5 & -\\
        4   &GMP 4570b      & tail      & 0.207 $\pm$ 0.026   & -                 &  108.8 & 3.9\\
        5	&GMP 4629a      & galaxy    & 0.066 $\pm$ 0.011   & -                 &  59.5 & -\\
        5   &GMP 4629b      & tail	    & 0.023 $\pm$ 0.007   & -                 &  26.9 & 2.8\\
        6	&D100a          & galaxy    & 1.099 $\pm$ 0.019   & $-$0.64 $\pm$ 0.15    &  113.9 & -\\
        6  &D100b          & tail      & 0.154 $\pm$ 0.024   & -                 &     179.2 & 8.1 \\
        7	&KUG1258+279Aa  & galaxy    & 2.845 $\pm$ 0.033   & $-$1.25 $\pm$ 0.12    &  134.4 & -\\
        7  &KUG1258+279Ab  & tail      & 1.070 $\pm$ 0.048   & $-$1.58 $\pm$ 0.51    &   273.3 & 9.8 \\ 
        8  &MRK58a         & galaxy    & 2.585 $\pm$ 0.058   & $-$2.13 $\pm$ 0.31    &   173.4 & -\\     
        8  &MRK58b         & tail      & 0.486 $\pm$ 0.058   & -                 &   173.4 & 5.8\\
        9  &IC 4040a       & galaxy    & 14.105 $\pm$ 0.036  & $-$1.22 $\pm$ 0.02    &   228.5 & -\\
        9  &IC 4040b       & tail      & 2.507 $\pm$ 0.059   & $-$1.41 $\pm$ 0.23    &   589.4 & 9.1\\
        10  &KUG1257+278a   & galaxy    & 0.584 $\pm$ 0.040   & -                 &   101.1 & -\\
        10  &KUG1257+278b   & tail      & 0.200 $\pm$ 0.031   & -                 &   61.4 & 2.9\\
\hline
	\end{tabular}
    \begin{tablenotes}
    \item {\sl Note:}
The properties of radio continuum for each galaxy/tail region defined in the radio continuum maps of Fig.~\ref{figure:tails_continuum} are listed. Sky area refers to the solid angle for each galaxy/tail region. The length of tails is estimated with a cut off at 2-sigma.
    \end{tablenotes}
    \label{tab:tail_table}
\end{table*}

A complementary tool for studying stripped tails is the radio continuum emission. At 1.4 GHz, radio continuum emission is dominated by synchrotron radiation which is emitted by the relativistic electrons moving within a magnetic field.
Like the colder, denser (traced by HI) and hotter, more diffuse (traced by H$\alpha$ and X-ray) gas, the plasma containing relativistic electrons and magnetic fields is stripped by ram pressure \citep[e.g.,][]{Gavazzi1987A&A...186L...1G}.
\citet{Murphy2009ApJ...694.1435M} identified radio-deficit regions along the outer edge of 6 Virgo Cluster galaxies, revealing that relativistic electrons and magnetic fields on their leading edges had been removed by ram pressure.
Furthermore, the stripped relativistic electrons can be re-accelerated (rejuvenated) in the tail \citep{Pinzke2013MNRAS.435.1061P}, either by turbulence and ICM shocks \citep{Kang2011ApJ...734...18K}, or by new SNe.
At the same time, local core-collapse supernovae can contribute to the relativistic electrons since H${\rm II}$ regions (tracing massive stars) have been found in RPS tails \citep[e.g.][]{Sun2007ApJ...671..190S,Yagi2010AJ....140.1814Y}.
However, fewer tails have been detected in radio continuum than in H$\alpha$ and X-ray. For example, there are 17 stripped tails detected in H$\alpha$ in the Coma cluster \citep{Gavazzi2018A&A...618A.130G}, but only 2 of them have thus far been detected in radio continuum \citep{Miller2009AJ....137.4436M}.
Currently most of the stripped tails seen in late-type galaxies in radio continuum are short and are detected in nearby clusters, for example NGC~4522 \citep{Vollmer2004AJ....127.3375V}, NGC~4402 \citep{Crowl2005AJ....130...65C} and others in the Virgo cluster.
A few long RPS tails, such as CGCG~097-073, CGCG~097-079 and UGC~6697 have been reported in Abell~1367 \citep{Gavazzi1978A&A....69..355G,Gavazzi1985ApJ...294L..89G,Gavazzi1987A&A...186L...1G,Gavazzi1995A&A...304..325G}.

Why do RPS tails tend not to be detected in the radio continuum?
How common are radio continuum tails behind RPS galaxies? How does the radio continuum emission in tails correlate with emission in other bands? Are the observed radio continuum tails mainly related to star formation in the tails or due to relativistic electrons which were stripped from the galaxy by ram pressure?
To address these questions deep radio continuum data are needed, something which has become feasible with the new-generation wide-band correlators deployed on existing radio telescopes.

As ram pressure stripping and SF activity in the tails are believed to be more prominent
in high-pressure environments than in low-pressure environments \citep[e.g.,][]{Sun2010ApJ...708..946S,Tonnesen2011ApJ...731...98T,Poggianti2016AJ....151...78P,Poggianti2017ApJ...844...48P},
the Coma cluster, as the most massive cluster at $z<0.025$, is an ideal target for these studies.
Coma has the richest optical data among nearby massive clusters, already with a sample of over 20 late-type galaxies with one-sided SF or ionized gas tails \citep{Smith2010MNRAS.408.1417S,Yagi2010AJ....140.1814Y,Kenney2015AJ....150...59K,Gavazzi2018A&A...618A.130G}. There has been an increasing effort to obtain multi-wavelength observations of these galaxies. RPS tails in bands other than H$\alpha$ have been detected. A spectacular example is D100, with a narrow tail observed in  X-rays with {\em Chandra} \citep{Sanders2014MNRAS.439.1182S} as well as CO detected at sub-mm wavelengths \citep{Jachym2017ApJ...839..114J}, co-existing with the narrow H$\alpha$ tail \citep{Yagi2007ApJ...660.1209Y}.

The deepest HI data on the Coma cluster to date are those of \cite{Bravo-Alfaro2000AJ....119..580B} and \cite{Bravo-Alfaro2001A&A...379..347B}, obtained with the VLA in its C--configuration. The angular resolution of $\sim 30''$ is not sufficient, though, for detailed study of the HI features in the galaxies and some of the narrow H$\alpha$ tails (e.g., D100's with a width of $\sim 4''$).
The deepest radio 1.4\,GHz continuum data on the Coma cluster were presented in \cite{Miller2009AJ....137.4436M}, before the implementation of the far more powerful WIDAR correlator.
In this paper, we present new HI (with higher spatial resolution) and 1.4\,GHz continuum (with deeper sensitivity) data on 20 RPS galaxies in the Coma cluster.

We assume $H_0$ = 70 km s$^{-1}$ Mpc$^{-1}$, $\Omega_m=0.3$, and $\Omega_{\Lambda}= 0.7$. At the redshift of the Coma cluster ($z=0.0231$), D$_L$=100.7 Mpc and $1^{\prime\prime}=0.466$ kpc. 

\section{Observations and Data reduction}

To further study Coma galaxies at radio frequencies, we obtained 1.4\,GHz continuum and HI data with the NRAO\footnote{The National Radio Astronomy Observatory is a facility of the National Science Foundation operated under cooperative agreement by Associated Universities, Inc.} Karl G. Jansky Very Large Array (VLA) in its B-configuration in two fields centered at NGC~4848 and D100, respectively (Fig.~\ref{figure:coma}), from June 1st to 11th, 2016 (Table~\ref{tab:observation_table}, program code: SH0174, PI: Sun). 
The 1.4\,GHz continuum data were taken with two base-bands (A0/C0 and B0/D0) covering a frequency range from 0.9 GHz to 2.1 GHz; each base-band is constructed with 7 spectral windows of 128 MHz, and each spectral window is divided into 64 channels of 2 MHz. 
The HI spectral data were taken with two spectral windows (one for A0/C0 and the other for B0/D0) of 64 MHz covering a velocity range of 1000 to 11200 km\,s$^{-1}$. Each spectral window is divided into 3584 channels of 17.8 kHz (or 3.88 km\,s$^{-1}$ velocity resolution).

The VLA was in the B-configuration for all the observations.
3C286 was observed to calibrate the flux density scale and the bandpass.
J1310+3220 was observed for the calibration of antenna gains and phase. 
The data were calibrated and reduced with the CASA software; each field was calibrated separately. 
Beyond the standard CASA pipeline, we removed radio frequency interference (RFI) carefully with the tfcrop and rflag mode of the flagdata task in CASA. About 40\% of the data in the continuum band and 20\% of the data in the HI band are identified as affected by RFI. Then, phase self-calibration was applied to the continuum data as there are strong sources (${\rm 15-40\ mJy\ beam^{-1}}$) in the fields which left residuals after the standard calibration. Complex gain calibration solutions were obtained with a 60s integration time sampling over 1-3 cycles of phase self-calibration until no further significant improvement was obtained.
Amplitude self-calibration was tested but was not applied as this brought no further improvements.

\renewcommand{\thefigure}{\arabic{figure}}

\begin{figure*}
 \centerline{\includegraphics[width=0.9\textwidth]{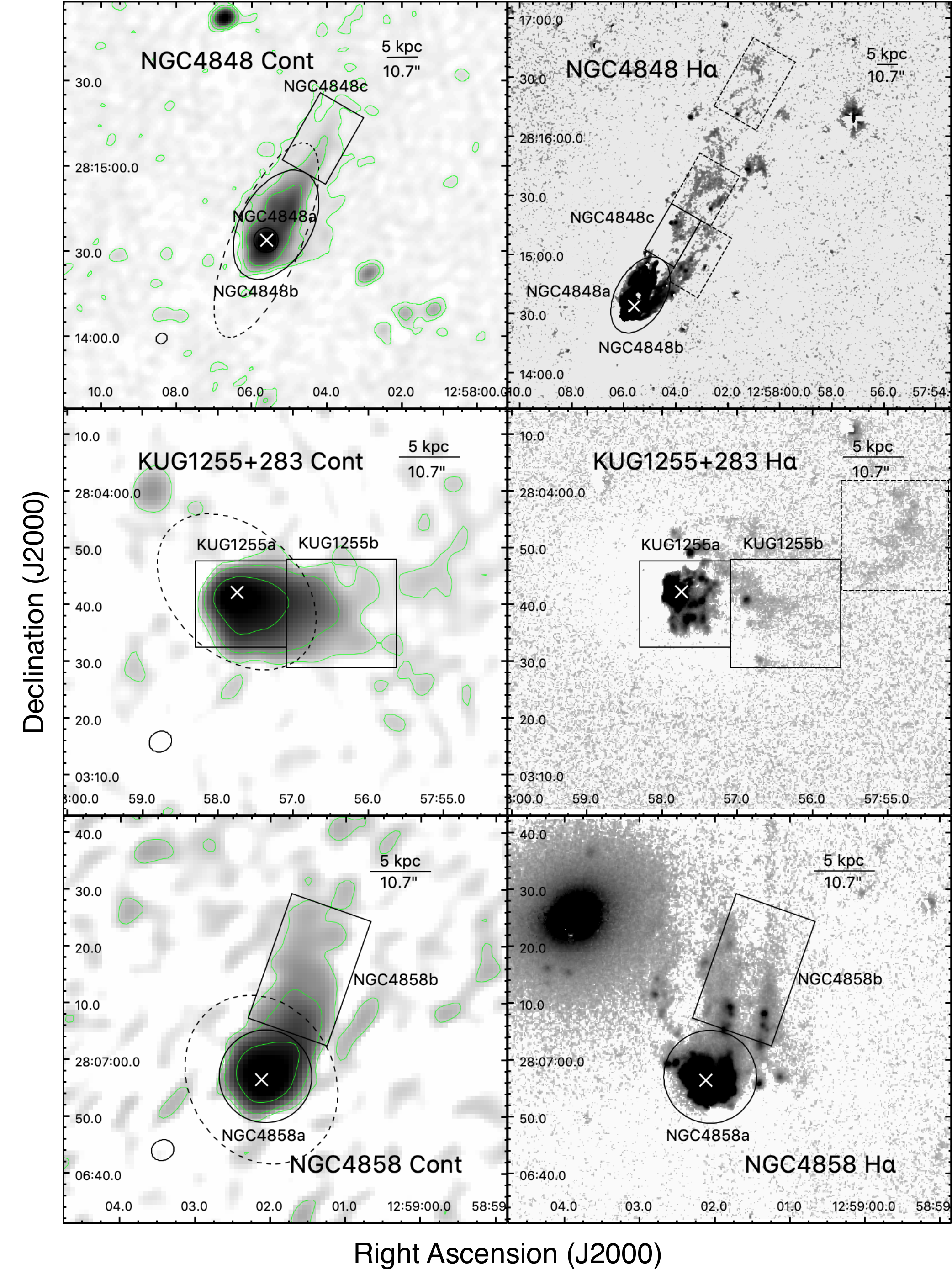}}  
 \vspace{-0.3cm}
 \caption{Left: VLA 1.4\,GHz continuum images for RPS galaxies.
Green contours show 2, 10 and 50 sigma flux density levels in the radio continuum.
Radio emission in galactic disks is measured in symmetrical regions around galaxy centers outlined by solid circles, or ellipses, or rectangles. Radio emission in one-sided asymmetric tails is also measured in regions shown by solid rectangles.
Dashed black ellipses show $D_{25}$.
Right: H$\alpha$ maps for the same galaxies from the {\em Subaru} data, except for KUG 1258+279A and NGC4921 (H$\alpha$ map from GOLDMine http://goldmine.mib.infn.it/). The nuclei are marked with a white X. The dashed boxes are where H$\alpha$ emission and upper limits on the radio continuum emission are measured, the values of which are included in Fig.~\ref{fig:cont-Ha}. The dashed line in the KUG 1258+279A H$\alpha$ map points along the direction that \citet{Smith2010MNRAS.408.1417S} identified as the `tail'.
}
 \label{figure:tails_continuum}
 \end{figure*}

\renewcommand{\thefigure}{\arabic{figure}}

\renewcommand{\thefigure}{\arabic{figure} (b)}
\addtocounter{figure}{-1}

\begin{figure*}
\centerline{\includegraphics[width=0.9\textwidth]{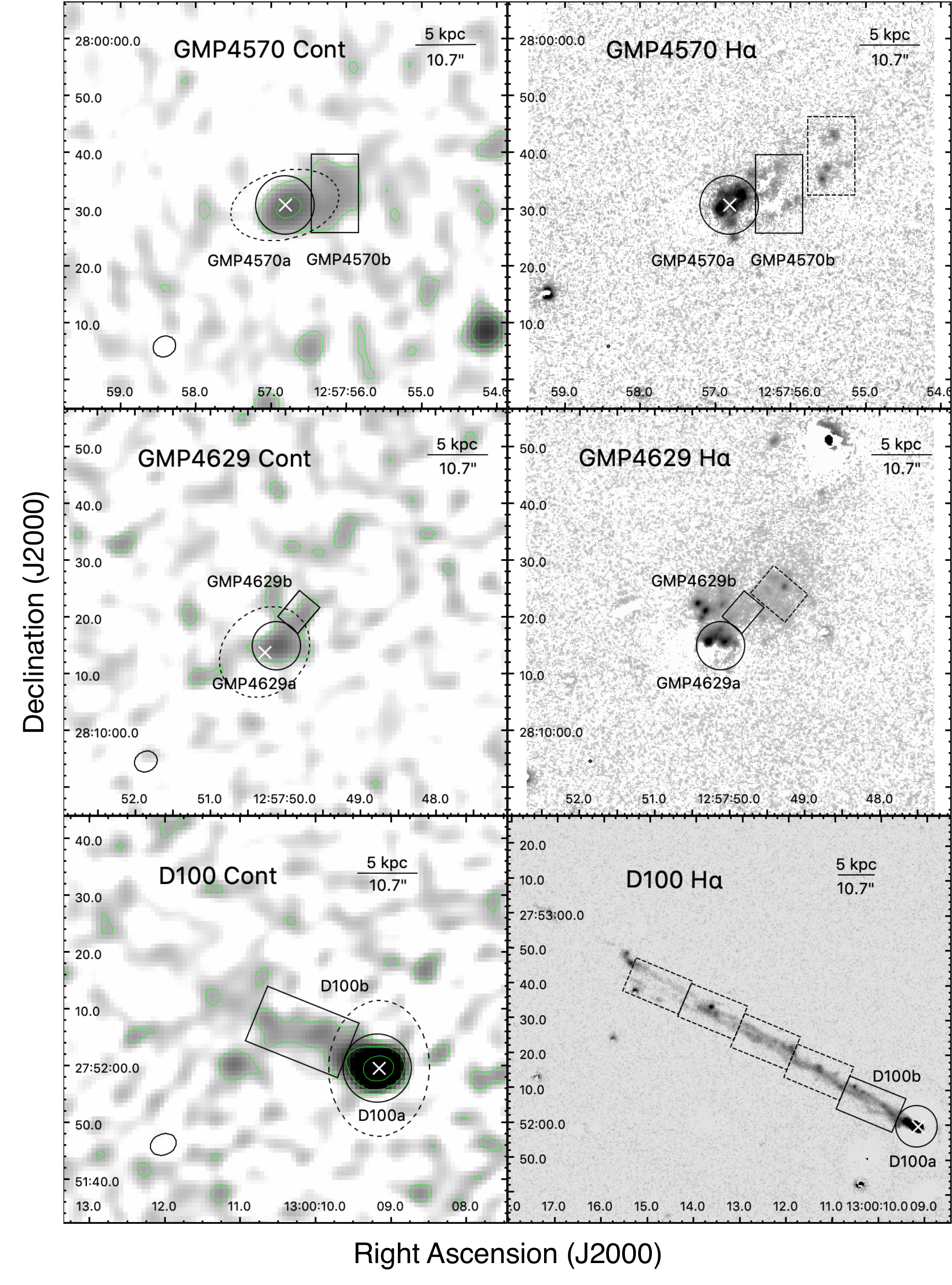}}  
 \vspace{-0.3cm}
\caption{}
\end{figure*}

\renewcommand{\thefigure}{\arabic{figure}}

\renewcommand{\thefigure}{\arabic{figure} (c)}
\addtocounter{figure}{-1}
 
\begin{figure*}
\centerline{\includegraphics[width=0.9\textwidth]{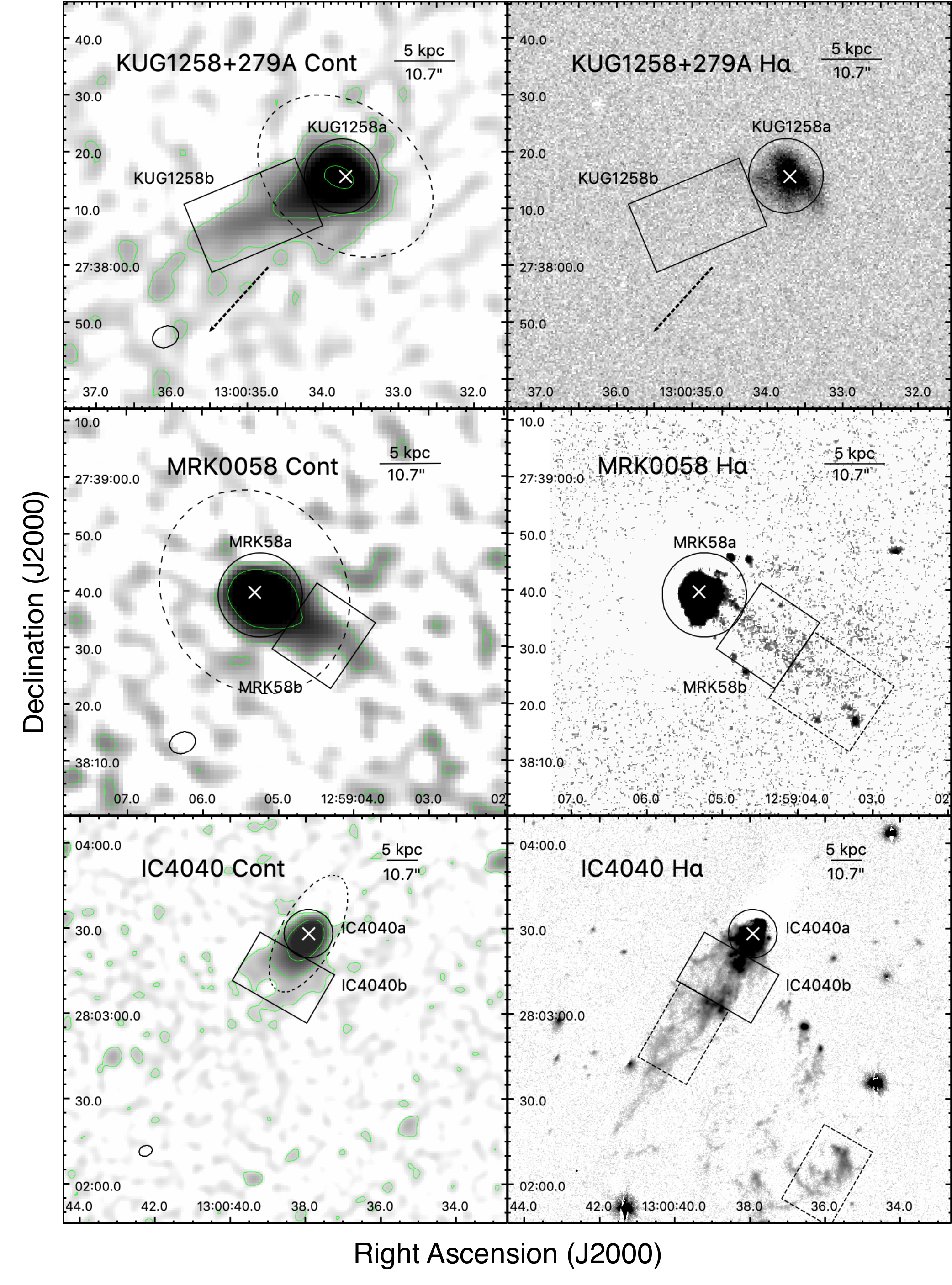}}
 \vspace{-0.3cm}
\caption{}
\end{figure*}

\renewcommand{\thefigure}{\arabic{figure}}

\renewcommand{\thefigure}{\arabic{figure} (d)}
\addtocounter{figure}{-1}
 
\begin{figure*}
\centerline{\includegraphics[width=0.9\textwidth]{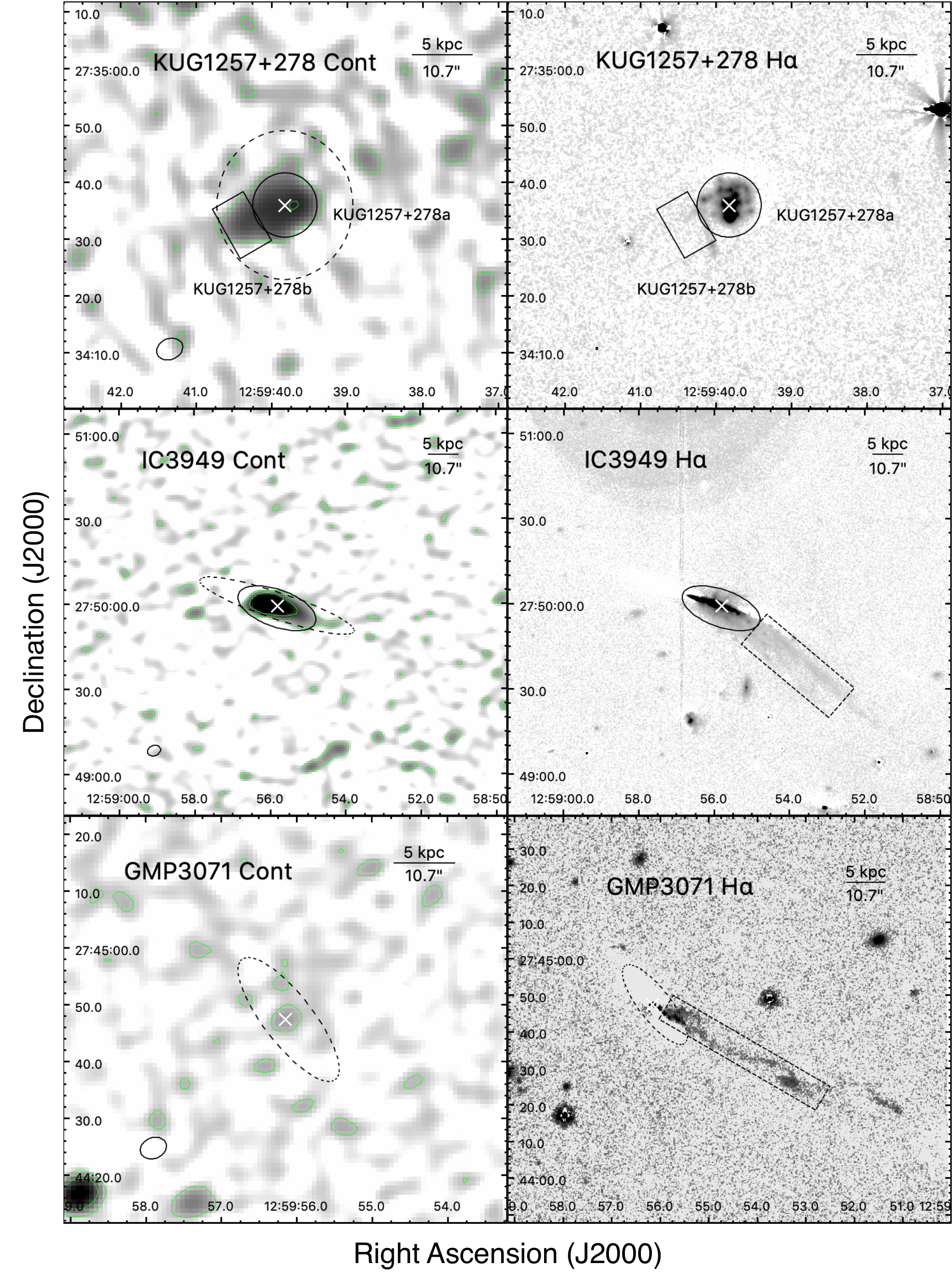}}
 \vspace{-0.3cm}
\caption{}
\end{figure*}

\renewcommand{\thefigure}{\arabic{figure}}

\renewcommand{\thefigure}{\arabic{figure} (e)}
\addtocounter{figure}{-1}
 
\begin{figure*}
\centerline{\includegraphics[width=0.9\textwidth]{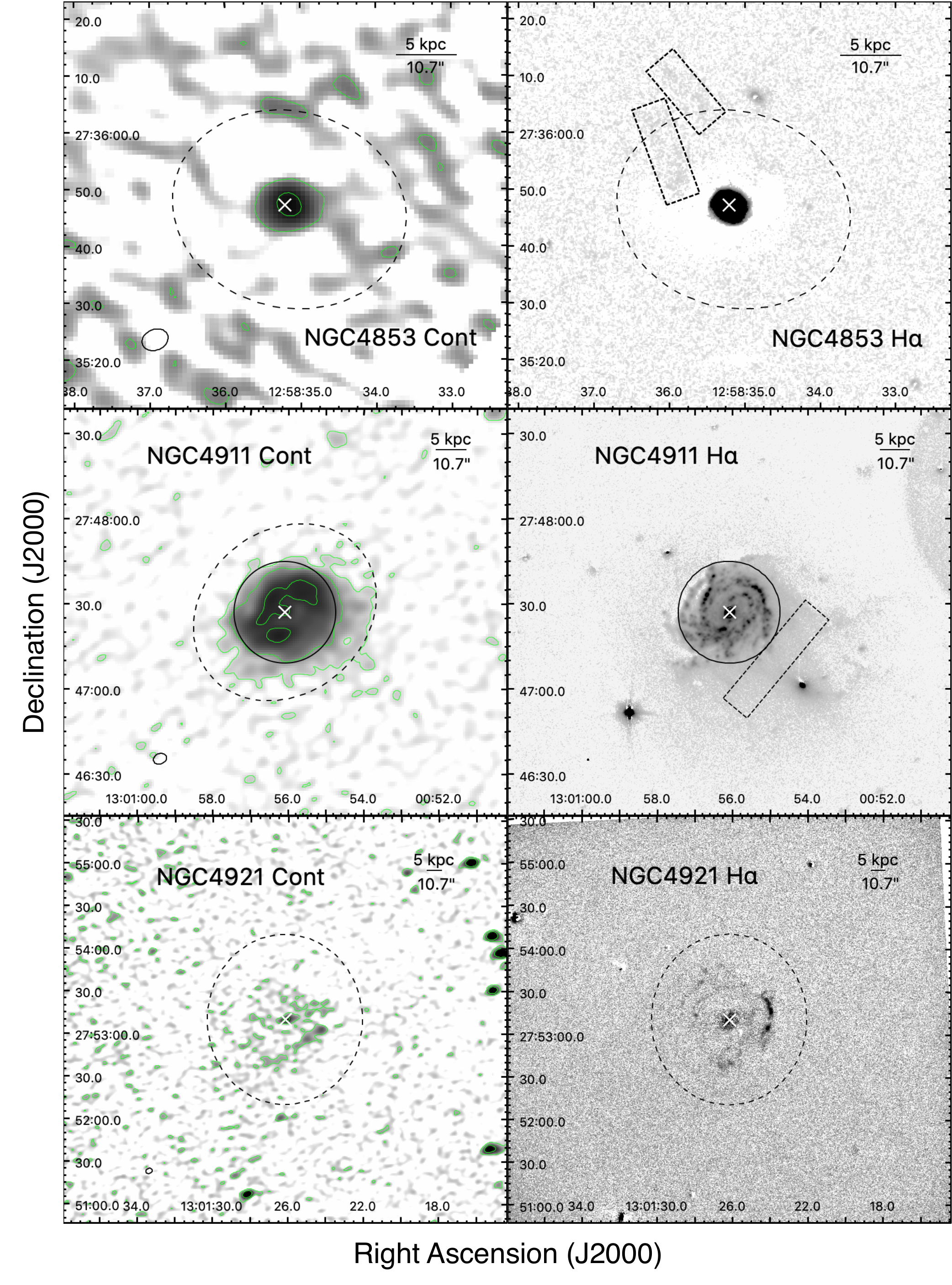}}
 \vspace{-0.3cm}
\caption{}
\end{figure*}

\renewcommand{\thefigure}{\arabic{figure}}

To create the HI cube, continuum was carefully fitted with line free channels in the range of $\pm 1000\ \rm{km\ s^{-1}}$ centred on the systemic velocity of each RPS galaxy; the continuum was subtracted with the uvcontsub algorithm.  
Multi-scale, multi-frequency, multi-term synthesis with w-projection (which is a wide-field imaging technique) were used in the tclean algorithm to clean the continuum and HI data.
The continuum map used a robust weighting of 0.5 to get a good compromise between optimized spatial resolution and sensitivity. 
The HI cube used  natural weighting to get the best sensitivity.

\section{Results and Analysis}
\label{sec:results} 

\subsection{1.4 GHz continuum sources}

The overall radio continuum map of the two observed fields
is shown in Appendix A.
The PyBDSF\footnote{http://www.astron.nl/citt/pybdsm/} source-detection package \citep{Mohan2015ascl.soft02007M} was used to locate and measure the flux densities of the radio sources from our data.
PyBDSF calculated the local rms for each pixel, evaluated over a 5\arcmin\ box. Then, source peaks greater than 4$\sigma$ are identified and all contiguous pixels higher than 2$\sigma$ are identified as belonging to one source. Finally, Gaussian fitting is used to resolve the source position and flux density.
In total, 1975 and 1173 radio sources (64 of them are duplicates) are detected within the 10\% response of  field 1 and 2, respectively.
Our source number density is 13.8 (for field 1) and 3.9 (for field 2) times of that from \citet{Miller2009AJ....137.4436M} in the central 6.4$\arcmin$ (90\% response radius of the VLA field at 1.4\,GHz).
To assess the reliability of our flux density measurement, we compared our results with those from 
\citet[][]{Miller2009AJ....137.4436M} and the FIRST survey \citep{White1997ApJ...475..479W} (see Appendix B for details). 
Our flux density is consistent with both.

With PyBDSF, we also derive the rms distribution for each field.
The rms in the center of fields 1 and 2 are 6 ${\rm \mu Jy}$ and 8 ${\rm\mu Jy}$ which are 4 and 3 times deeper than the \citet{Miller2009AJ....137.4436M} data. Because of the primary beam correction, the rms increases from the center of the field to the outer regions. As NGC~4848 and D100 are at the center of their fields, they have the deepest data in their respective fields.
The full continuum source catalog is shown in Table~\ref{tab:catolog_table}.

\begin{figure}
    \hspace{-0.15cm}
	\centerline{\includegraphics[width=1.05\columnwidth]{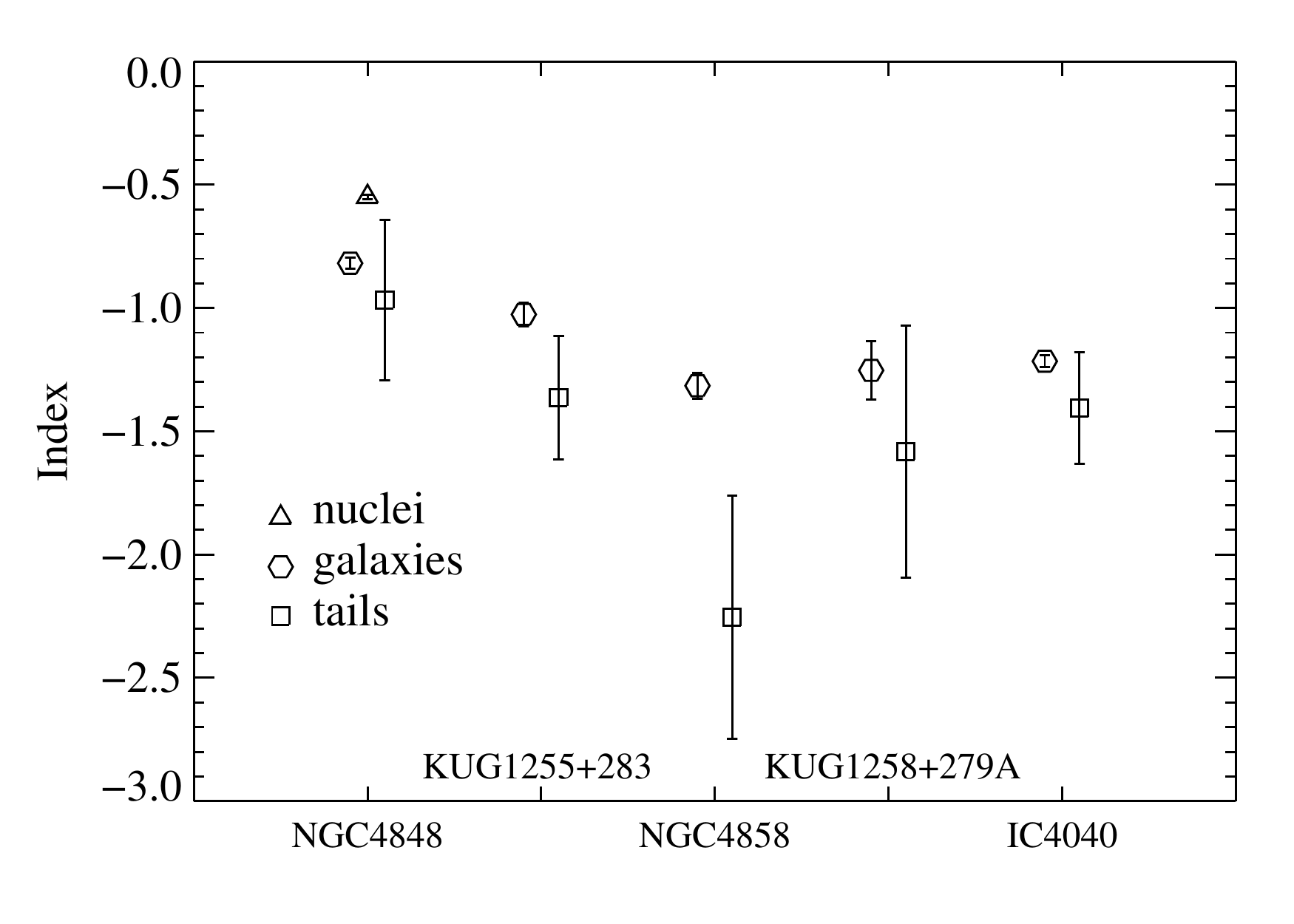}}
	\vspace{-0.5cm}
    \caption{The in--band spectral index for the nucleus, galaxy and tail regions of five RPS galaxies with the spectral index in the tail constrained.
    }
    \label{fig:Index_map_figure}
\end{figure}

\subsection{Radio continuum emission of RPS galaxies}

Twenty RPS galaxies \citep{Smith2010MNRAS.408.1417S,Yagi2010AJ....140.1814Y,Kenney2015AJ....150...59K} are covered by our observations. Their radio properties, measured from our data, are listed in Table~\ref{tab:tail_sources_table}.
15 of them are detected in the radio continuum, while the other five (GMP~4060, NGC~4854, GMP 3016, GMP 2923 and KUG 1258+277) were not detected (Table~\ref{tab:tail_sources_table}).

We detected significant extended radio emission coincident with the H$\alpha$ or SF tails in 10 of the 20 RPS galaxies (Fig.~\ref{figure:tails_continuum} \& Table~\ref{tab:tail_table}), or in 10 of 15 (66\%) RPS galaxies detected at 1.4\,GHz, unambiguously revealing the widespread occurrence of relativistic electrons and magnetic fields in the stripped tails.
Radio tails behind IC~4040 and KUG~1255+283 were reported in \citet{Miller2009AJ....137.4436M}, while the other 8 are new.
The galaxy regions and tail regions are visually defined based on the radio continuum maps (Fig.~\ref{figure:tails_continuum}).
Symmetrical disks are defined as galaxy regions from radio continuum map by means of circles/ellipses/rectangles. The one-sided asymmetric structures extending beyond the galaxy regions are defined as tails (solid rectangles).
Both galaxy regions and tail regions were set to match approximately the 2-sigma contour of the radio emission.
The 1.4 GHz continuum flux densities of galaxies and tails are listed in Table~\ref{tab:tail_table}. 

The radio continuum tail is always spatially coincident with the H$\alpha$ or SF tail but usually shorter than either (except for KUG 1257+278), at least at the sensitivity level of the current radio data.
All tails, except for KUG 1257+278, extend beyond $D_{25}$ (diameter of a galaxy at an isophotal level of 25 mag\,arcsec$^{-2}$ in the $B$-band, taken from LEDA \citep{Makarov2014A&A...570A..13M}.)

The in-band spectral index was also derived, taking advantage of the wide bandwidth covered by the correlator. Because of the weak signal in general, we divided the 1.4\,GHz continuum into two parts, a low frequency  (approximately 0.9 --- 1.5 GHz) and a high frequency one (approximately 1.5 --- 2.1 GHz). Then flux density maps centered at 1.25 GHz and 1.75 GHz were created, and the power law index for the galaxy and tail was derived using the convention of S $\sim$ $\nu^\alpha$.
We only measure the spectral index of a region when it is detected above 5$\sigma$ in both bands. The results are shown in Table~\ref{tab:tail_table} and Fig.~\ref{fig:Index_map_figure}.
Our results hint that the tail has a steeper spectrum than the galaxy, which is consistent with aging of electrons and a lack of fresh injection of relativistic electrons in the tails. But the limited depth and frequency coverage of our data do not allow us to reach a definite conclusion.

\begin{figure}
	\centerline{\includegraphics[width=0.88\columnwidth]{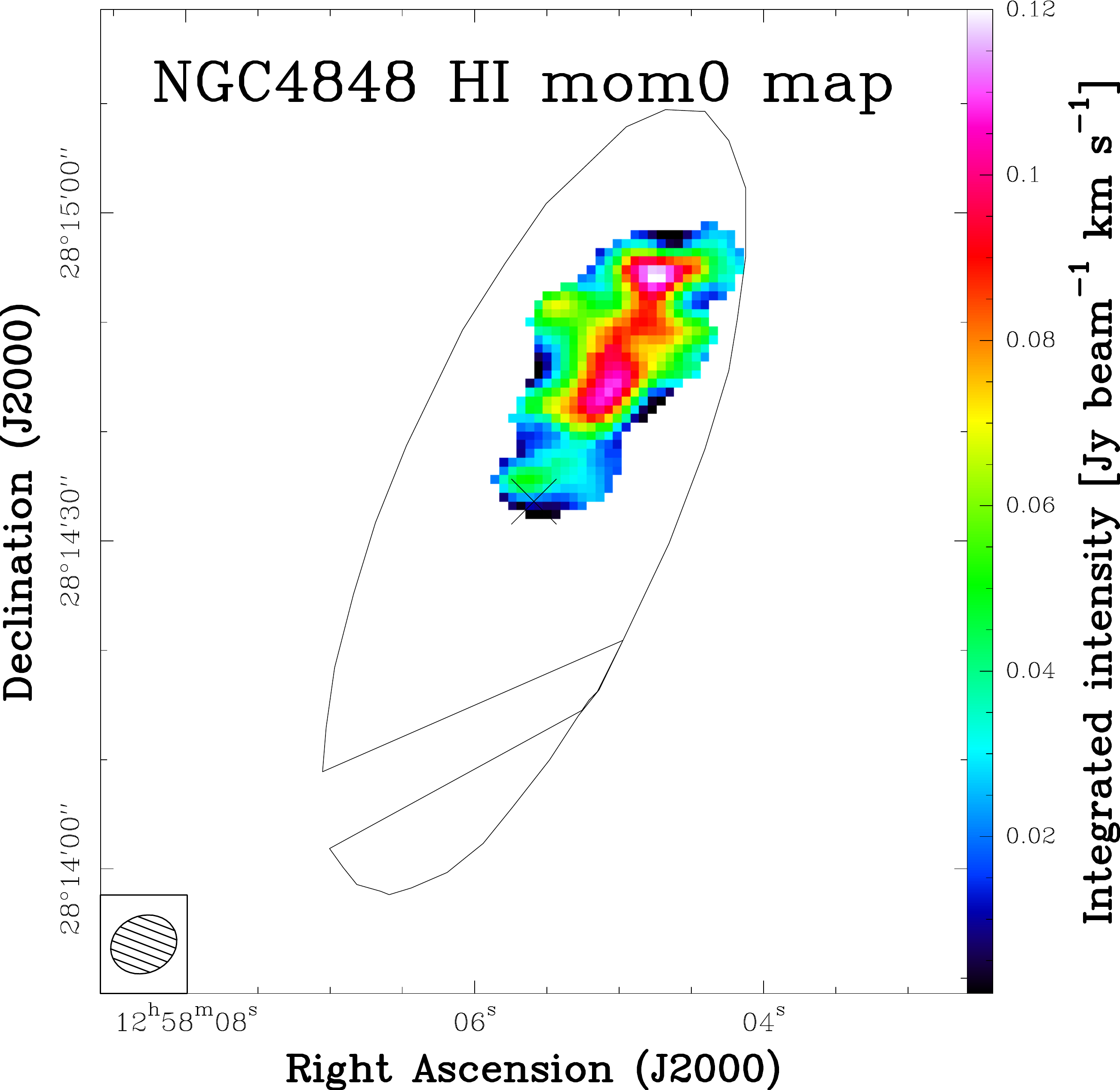}}
    \centerline{\includegraphics[width=0.88\columnwidth]{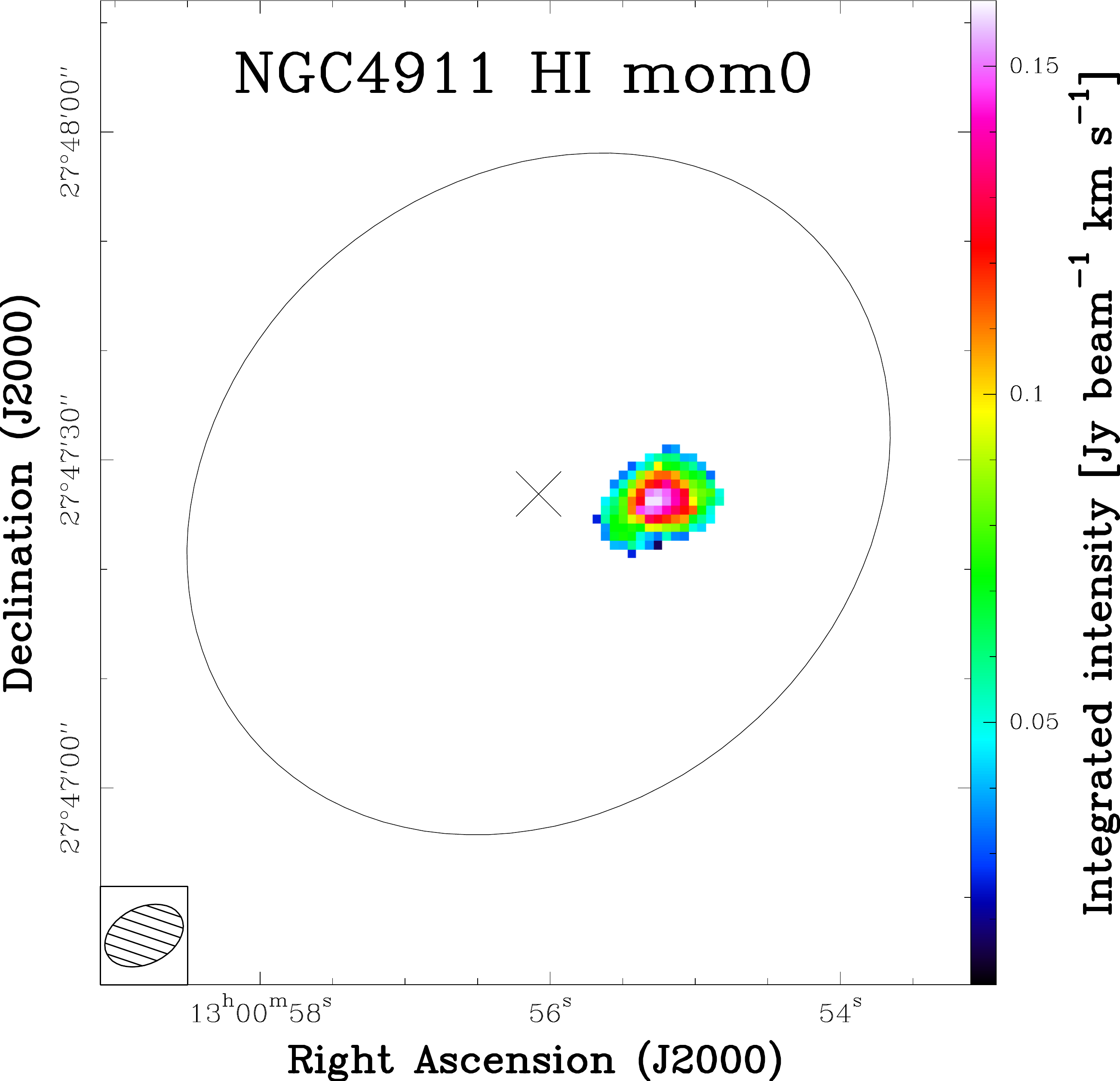}}
    \centerline{\includegraphics[width=0.88\columnwidth]{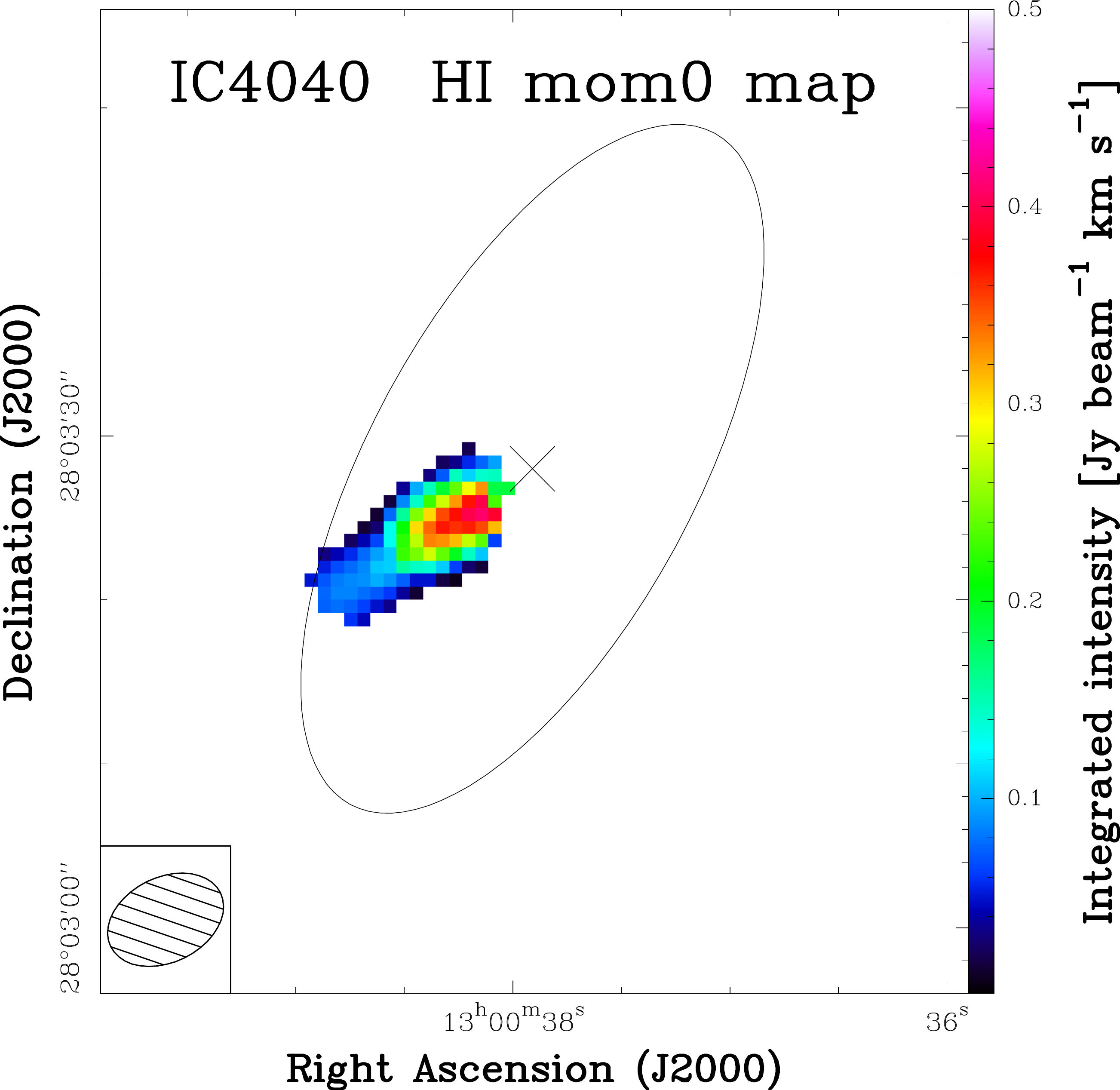}}
    \caption{Integrated intensity (moment 0) map of HI within the 2$\sigma$ 3D mask created by SoFiA. The beam size is listed in Table~\ref{tab:observation_table}. Black ellipses show the $D_{25}$ regions. The nuclei of galaxies are shown as black X.}
    \label{fig:HI_moment0_map_figure}
\end{figure}

\begin{figure}
    \centerline{\includegraphics[width=0.96\columnwidth]{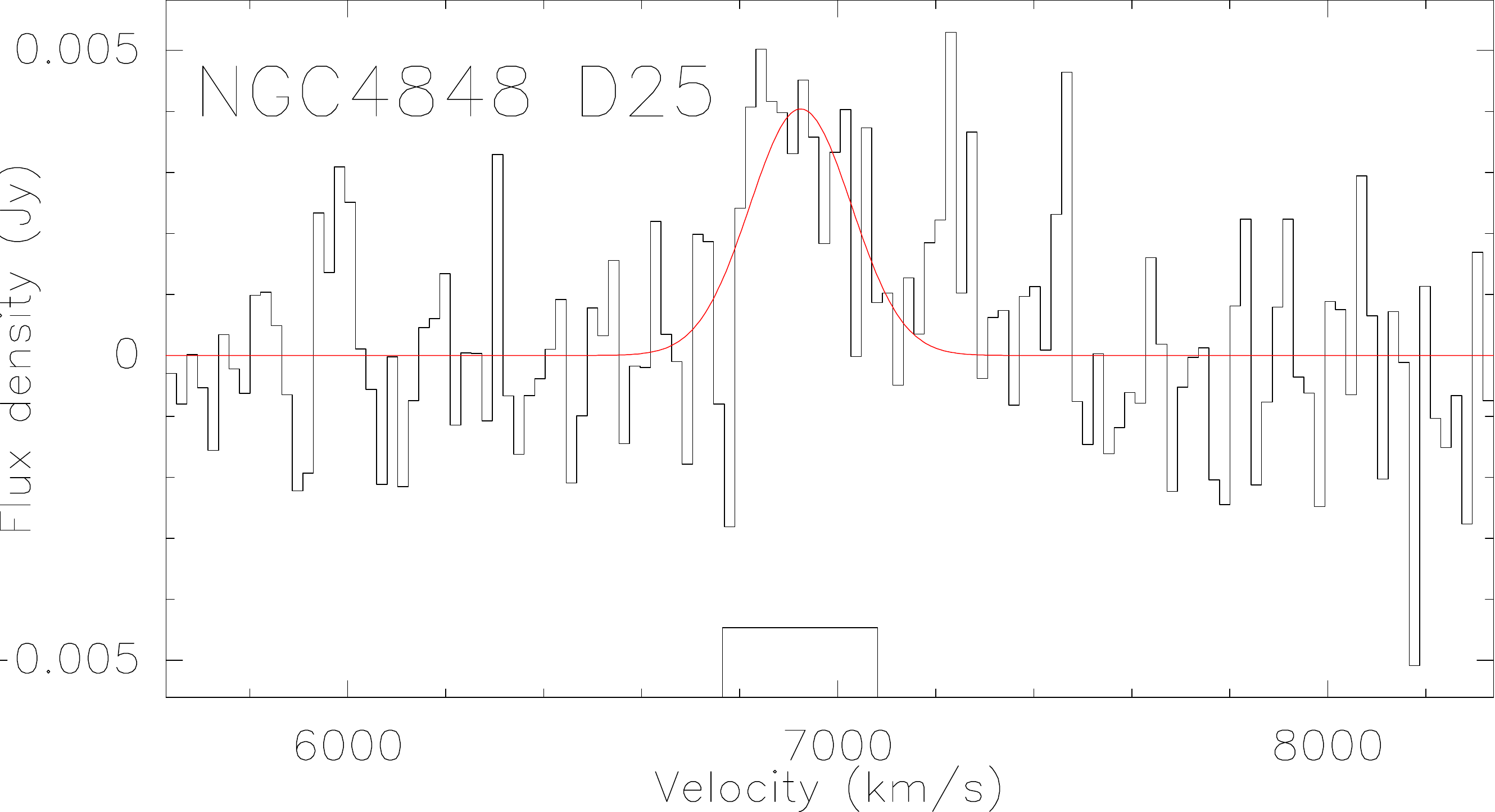}}
    \centerline{\includegraphics[width=0.96\columnwidth]{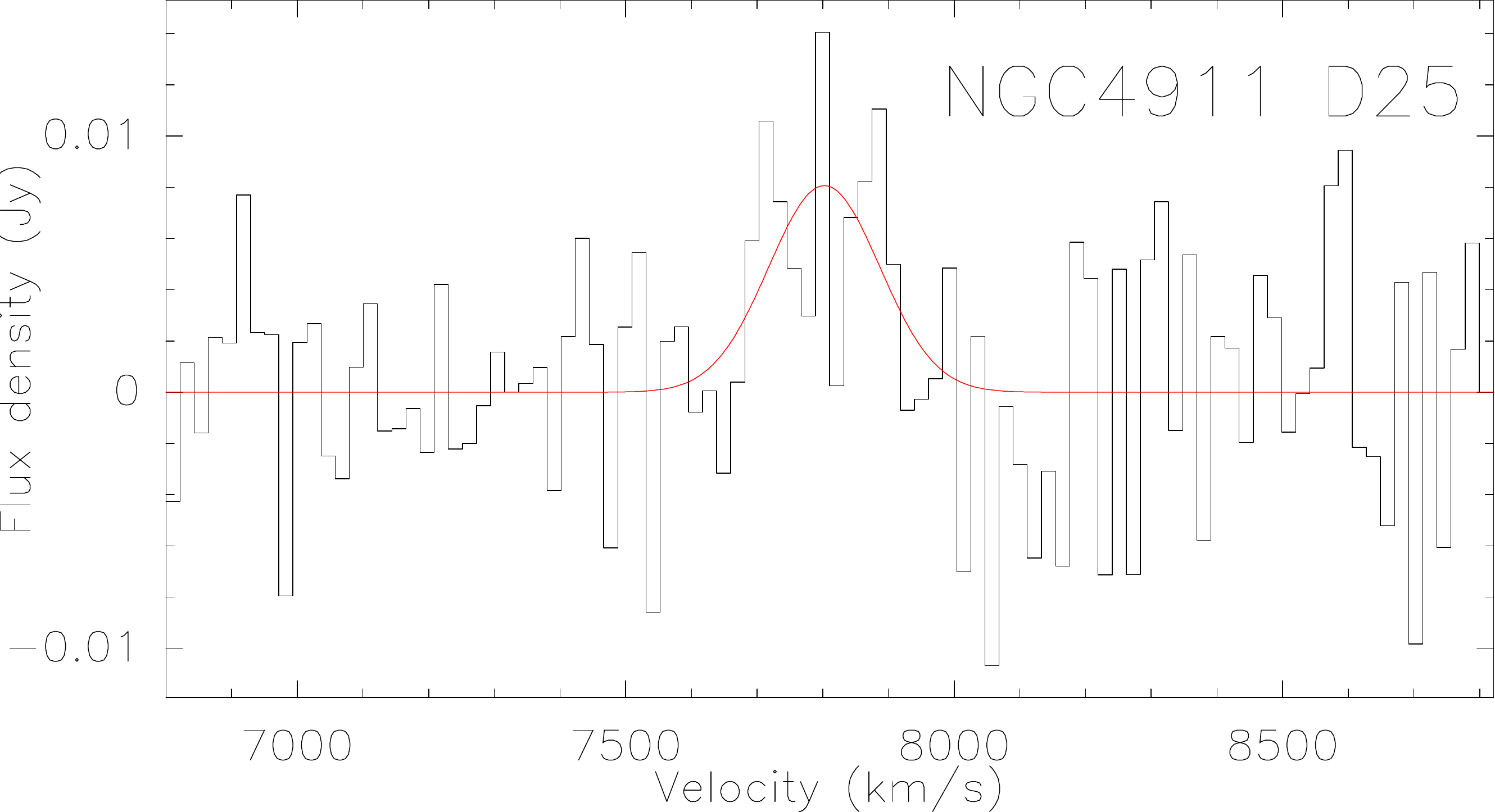}}
    \centerline{\includegraphics[width=0.96\columnwidth]{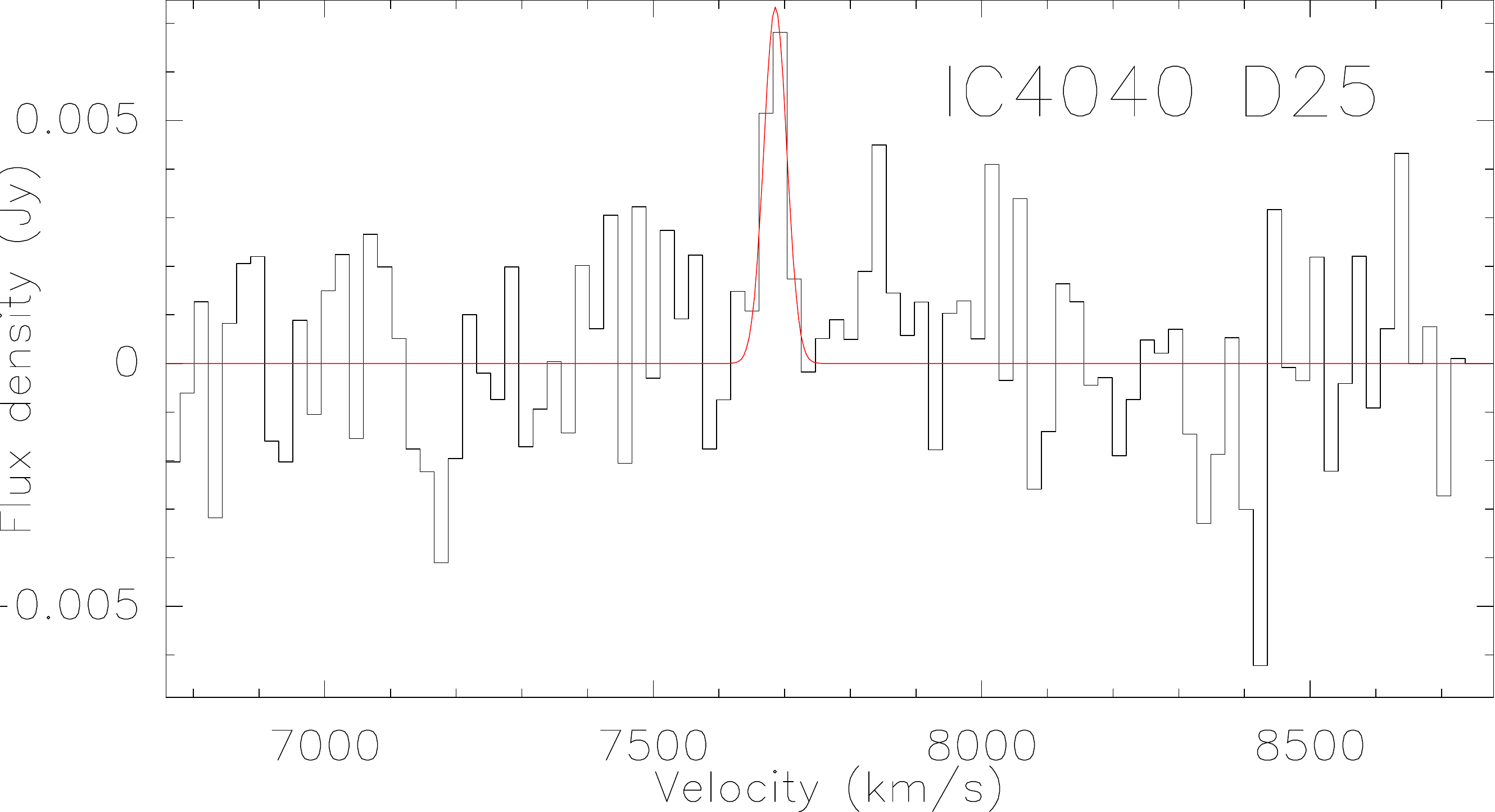}}
    \caption{HI spectra for the $D_{25}$ regions of three galaxies with HI detected. The HI spectra are fitted with a Gaussian function shown as the red line.}
    \label{fig:HI_spectra_figure}
\end{figure}

\begin{table*}
	\caption{HI deficiency for late-type galaxies in the Coma cluster}
	\begin{tabular}{cccccccccccc} 
		\hline
		NO. & Galaxy & Type & $D_{25}$ & $M_{\rm HI}$${\rm _{normal}}$  & rms$_{\rm HI}$ & $m_{\rm B}$ & $i$ & $M_{\rm HI}$ & Def$_{\rm HI}$ & Def$_{\rm HI\_BA00}$ \\
            &              &                    & [arcsec] & ${\rm [10^8 M_\odot]}$&   [${\rm \mu Jy\ beam^{-1}}$]  & [mag] & [degree] & ${\rm [10^8 M_\odot]}$ & \\ 
		\hline
		1	&NGC~4848 (GMP~4471)     & Scd & 73.8 & 87.7  & 149    & 13.43& 73.9	    & 19.99     & 0.64     & 1.14 \\
        2	&KUG~1255+283 (GMP~4555) & I   & 32.2 & 16.7  & 284  & 15.55	& 47.4 & $<5.58$ &$>0.48$  & -\\    
        3	&NGC~4858 (GMP~3816)     & Sbc & 30.8 & 14.6  & 362  & 15.42	& 35.6 & $<6.77$ &$>0.33$  & $>0.93$\\
        4	&GMP~4570 	  & I   & 19.4 & 6.1   & 395  & 16.74	& 59.5 & $<4.05$ & $>0.18$  & - \\
        5	&GMP~4629     & N/A & 17.3 & 4.2   & 207  & 17.13	& 37.8 & $<1.77$ &$>0.38$  & - \\
        6	&D100 (MRK~0060, GMP~2910) & I & 23.9 & 9.2   & 246  & 15.83	& 45.0 & $<3.20$ & $>0.46$  & -\\       
        7	&KUG~1258+279A (GMP~2599) & Sb& 33.7 & 16.7  & 440  & 15.32	& 44.6 & $<8.64$ &$>0.29$ & 0.58\\
        8	&MRK~0058 (GMP~3779)     & Sb  & 37.0 & 20.1  & 788  & 14.98	& 31.1 & $<16.61$ & $>0.08$  & 0.51\\
        9	&IC~4040 (GMP~2559)      & Scd & 45.5 & 33.3  & 388    & 14.60 & 83.6 & 7.5     & 0.65  & 0.61\\       
        10	&KUG~1257+278 (GMP~3271) & I   & 26.2 & 11.0  & 697  & 16.34	& 24.0 & $<7.90$ & $>0.15$  & -\\       
        11	&IC~3949 (GMP~3896)      & S   & 57.3 & 46.0  & 533  & 14.91	& 90.0 & $<10.46$ & $>0.64$  & - \\
        12	&GMP~3071     & S0/a& 26.8 & 6.4   & 290  & 16.94	& 90.0 & $<2.84$ & $>0.35$  & - \\
        13	&NGC~4853 (GMP~4156)    & S0p & 41.5 & 15.2  & 1649  & 14.21 & 40.3	& $<47.27$ & $>-0.49$ & -\\
        14	&NGC~4911 (GMP~2374)     & Sb  & 68.9 & 69.7  & 326    & 13.44 & 34.7 & 37.34     & 0.27     & 0.58\\ 
        15	&NGC~4921 (GMP~2059)     & Sb  & 119.7& 210.4 & 504  & 13.34	& 24.8 & $<39.85$ & $>0.72$  & 1.11\\       
        16	&GMP~4060     & Sd & 18.1& 5.3  & 849  & 17.36	& 34.4 & $<6.55$ &$>-0.09$ & - \\
        17	&NGC~4854 (GMP~4017)    & S0  & 57.3 & 29.0  & 1019 & 14.77	& 53.8 & $<38.35$ & $>-0.12$ & - \\       
        18	&GMP~3016      & N/A & N/A  & -   & 372  & N/A	    & N/A     & -     & - & -\\
        19	&GMP~2923     & E & 18.5& 3.0   & 267  & 17.32   & 90.0 & $<2.13$ & $>0.15$  & -\\       
        20	&KUG~1258+277 (GMP~2640)   & S0p& 29.4& 7.6   & 965  & 15.62   & 73.4 & $<13.19$ &$>-0.24$ & $>0.7$ \\                
		\hline
	\end{tabular}
    \begin{tablenotes}
    \item {\sl Note:}
The expected HI mass ($M_{\rm HI}$${\rm _{normal}}$) is derived from table IV of \citet{Haynes1984AJ.....89..758H} with $D_{25}$ from LEDA and galaxy type from \citet{Dressler1980ApJS...42..565D} (except for GMP~4060 and GMP~2923 from LEDA). For the galaxies without morphology type information (N/A), $M_{\rm HI}$${\rm _{normal}}$ is derived from the average relation of $M_{\rm HI}$ and $D_{25}$ for all the galaxies. The rms$_{\rm HI}$ values are calculated for the HI cubes for a channel width of 21.5 km\,s$^{-1}$. The beam sizes of the HI cubes are shown in Table~\ref{tab:observation_table}. The values for $m_{\rm B}$ and inclination ($i$) are also from LEDA. N/A means not available. Def$_{\rm HI\_BA00}$ is based on Def$_{\rm HI}$ from \citet{Bravo-Alfaro2000AJ....119..580B}.
\end{tablenotes}
     \label{tab:HI_Deficiency}
\end{table*}

\begin{table*}
	\caption{HI flux of RPS galaxies}
	\begin{tabular}{cccc} 
		\hline
		 Galaxy &HI Flux$_{\rm SoFiA}$ & HI Flux$_{\rm D25}$ & HI Flux$_{\rm BA00}$\\%
                      &  ${\rm [Jy\ km\ s^{-1}]}$ & ${\rm [Jy\ km\ s^{-1}]}$ & ${\rm [Jy\ km\ s^{-1}]}$ \\
		\hline
		NGC~4848 &$0.464\pm0.005$ & $0.878\pm0.132$ & $0.37\pm0.03$   \\
		NGC 4911 &$0.136\pm0.004$ & $1.64\pm0.32$ & $0.80\pm0.07$   \\
        IC 4040  &$0.260\pm0.005$ & $0.33\pm0.11$ & $0.29\pm0.06$  \\ 
\hline
	\end{tabular}
    \begin{tablenotes}
    \item {\sl Note:}
HI Flux$_{\rm SoFiA}$ is the HI flux calculated in the 2$\sigma$ 3D mask created by SoFiA. HI Flux$_{\rm D25}$ is calculated in the $D_{25}$ region defined in the optical (black ellipse in Fig.~\ref{fig:HI_moment0_map_figure}). HI Flux$_{\rm BA00}$ is measured by \citet{Bravo-Alfaro2000AJ....119..580B}.
    \end{tablenotes}
     \label{tab:HI_table}
\end{table*}

\subsection{HI emission of RPS galaxies}
The HI data have a spatial resolution of $\sim 6\arcsec$ using natural weighting to match the narrow width of the gaseous stripped tails of RPS galaxies.
The sensitivities of the HI spectra for 20 RPS galaxies are listed in Table~\ref{tab:HI_Deficiency}.
Although our high spatial resolution HI data are less sensitive than the low resolution data from \citet{Bravo-Alfaro2000AJ....119..580B} (observed in C configuration with the VLA) for sources with radii larger than 30$\arcsec$,
the better spatial resolution can resolve the peak HI structure of galaxies and give better HI upper limits for small galaxies and narrow tails, such as D100's tail which is discussed in section 4.2.2.  

HI emission is detected in the integrated intensity (moment 0) maps for NGC~4848, NGC 4911 and IC 4040 as shown in Fig.~\ref{fig:HI_moment0_map_figure}. The HI integrated intensity maps are made from the HI cubes with 2$\sigma$ 3D masks created by SoFiA \citep{Serra2015MNRAS.448.1922S}.
The $D_{25}$ regions are also indicated in Fig.~\ref{fig:HI_moment0_map_figure}. The gap in the $D_{25}$ region for NGC~4848 is the result of a mask which was applied to remove an image side-lobe artifact.
The positions of HI emission peaks in the integrated intensity maps are consistent with those from \citet{Bravo-Alfaro2000AJ....119..580B}.
Corresponding HI spectra for the $D_{25}$ regions are shown in Fig.~\ref{fig:HI_spectra_figure} and the HI results are listed in Table~\ref{tab:HI_table}.
The HI flux within the SoFiA 3D mask (HI Flux$_{\rm SoFiA}$) is lower than that within $D_{25}$ (HI Flux$_{\rm D25}$), revealing that there is some weak HI emission within $D_{25}$ that was not picked up by SoFiA, especially for NGC 4911. Increasing the radii of the areas of integration to twice $D_{25}$ does not lead to a further increase in HI flux, suggesting that we have measured the total flux within $D_{25}$. Our HI Flux$_{\rm D25}$ is larger than HI Flux$_{\rm BA00}$; these may be because \citet{Bravo-Alfaro2000AJ....119..580B} ignored the diffuse HI emission within $D_{25}$ where the S/N ratio is lower than 3. 

Six galaxies (NGC~4848, MRK 0058, KUG 1258+279A, IC 4040, NGC 4911 and NGC 4921) in our sample are detected in HI by \citet{Bravo-Alfaro2000AJ....119..580B},  three of which (MRK 0058, KUG 1258+279A and NGC 4921) are not detected in this work. The HI in these three galaxies is probably of an extended nature and lacks bright clumps. Also, these galaxies fall at about the 50\% response radius of field 2 so the sensitivity of the data may not be good enough.

\section{Discussion}

\subsection{The nature of the radio continuum emission of the stripped tails}

The observed 1.4\,GHz continuum should be dominated by synchrotron radiation. Typically for spiral galaxies, the radio continuum emission is a tracer for star formation as relativistic electrons are accelerated by core-collapse supernovae \citep[e.g.,][]{Condon1991ApJ...376...95C,Bell2003ApJ...586..794B,Murphy2008ApJ...678..828M}.
As H${\rm II}$ regions (tracing massive stars) have been found in stripped tails \citep{Gavazzi1995A&A...304..325G,Sun2007ApJ...671..190S,Cramer2019ApJ...870...63C}, the local supernovae could contribute to the synchrotron radiation in tails.
On the other hand, relativistic electrons can be stripped by ram pressure as in radio head-tail galaxies \citep[e.g.,][]{Gavazzi1987A&A...186L...1G,Murphy2009ApJ...694.1435M}. 
In the following we investigate if the observed radio continuum emission is mainly related to star formation in the tail or due to relativistic electrons which were stripped from the galaxy by ram pressure.

\subsubsection{Radio continuum versus H$\alpha$ and IR}

\begin{figure}
	\centerline{\includegraphics[width=1.13\columnwidth]{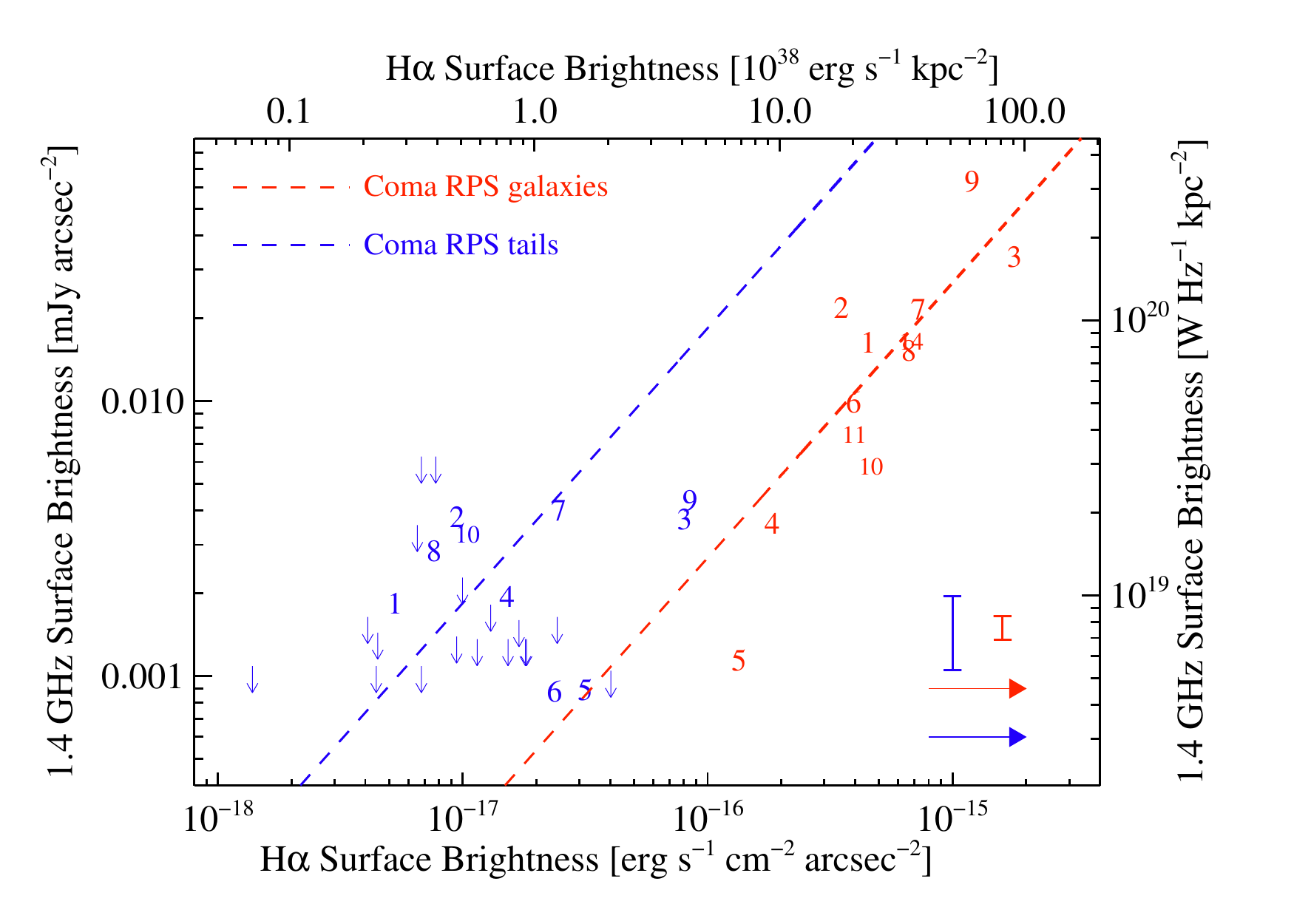}}
    \caption{The 1.4 GHz surface brightness versus H$\alpha$ surface brightness for RPS galaxies and their tails in regions defined in Fig.~\ref{figure:tails_continuum}. Each number indicates a galaxy as listed in Table~\ref{tab:tail_sources_table}, and tails are in blue and galaxies are in red. Blue arrows are 1.4 GHz upper limits for the dashed box regions in Fig.~\ref{figure:tails_continuum} where an H$\alpha$ tail is detected but a radio extension is absent from the current data. Bottom right arrows show the extinction correction of one magnitude applied on H$\alpha$.
    The mean ratios of 1.4 GHz to H$\alpha$ for RPS tails and galaxies in the Coma cluster are shown as blue and red dashed lines respectively. 
    }
    \label{fig:cont-Ha}
\end{figure}

We compared the 1.4 GHz continuum and H$\alpha$ surface brightness (SB) in tails (blue) and galaxies (red) in Fig.~\ref{fig:cont-Ha}, to study the origin of the radio tails. Each number indicates a galaxy as listed in Table~\ref{tab:tail_sources_table}.
Blue arrows show 3 sigma 1.4 GHz SB upper limits for the dashed box regions in Fig.~\ref{figure:tails_continuum}.
The 1.4 GHz SB is the ratio between 1.4 GHz flux density and sky area (listed in Table~\ref{tab:tail_table}) for each tail/galaxy region defined in Fig.~\ref{figure:tails_continuum}.
H$\alpha$ SB values for the same tail/galaxy regions are estimated from the {\em Subaru} data as in \citet{Yagi2017ApJ...839...65Y}, except that of KUG~1258+279A which is from GOLDMine.
Broadband over--subtraction and [NII], [SII], and [OI] contamination are corrected.
The intrinsic extinction is unknown.  About 1 mag of intrinsic extinction was measured for two isolated HII regions in the Virgo cluster \citep{Gerhard2002ApJ...580L.121G,Cortese2004A&A...416..119C}. We simply adopt this value on our H$\alpha$ SB.

If the relativistic electrons emitting radio in stripped tails come from local supernovae in the tails, it is expected that the same 1.4 GHz --- H$\alpha$ relation in galaxies will be followed. 
However, all the RPS galaxies, except IC~4040 (9 in Fig.~\ref{fig:cont-Ha}), show a higher 1.4 GHz to H$\alpha$ ratio in the tails than in the galaxies. The mean ratio in the tails ($1.84\times10^{14}$ mJy\,erg$^{-1}$\,s\,cm$^2$, blue dashed line) is 6.8 times that in galaxies ($2.69\times10^{13}$ mJy\,erg$^{-1}$\,s\,cm$^2$, red dashed line).
Moreover, H$\alpha$ emission in tails only provides an upper limit on the SFR as the stripping and heating (caused by the RPS shock) of ionized gas can account for much of the diffuse H$\alpha$ emission \citep[e.g.,][]{Sun2010ApJ...708..946S,Fossati2016MNRAS.455.2028F,Cramer2019ApJ...870...63C}. 
So the radio continuum in the Coma tails is generally too strong compared to the value expected from the SFR.
Our results suggest that the radio-emitting plasma in tails is not formed in situ, but stripped from the galaxy by ram pressure.
While the above analysis is based on rectangular regions roughly matching the radio emission, we also examined the change on the results by using tail regions coinciding with a constant surface brightness limit of 20 $\mu$Jy\ beam$^{-1}$ (beam size $4.7\arcsec \times 3.7\arcsec$). Both radio surface brightness and the H$\alpha$ surface brightness have been re-measured with this new set of regions. 
The averaged 1.4 GHz to H$\alpha$ ratio in the new set of the tail regions is now 46\% higher than the blue dashed line in Fig.~\ref{fig:cont-Ha}, and the scatter still remains about the same. 
For simplicity and the sizable scatter here, we stick to the results from the simple rectangular regions as shown in Fig.~\ref{figure:tails_continuum}.

\begin{figure}
    \vspace{-0.7cm}
	\centerline{\includegraphics[width=1.13\columnwidth]{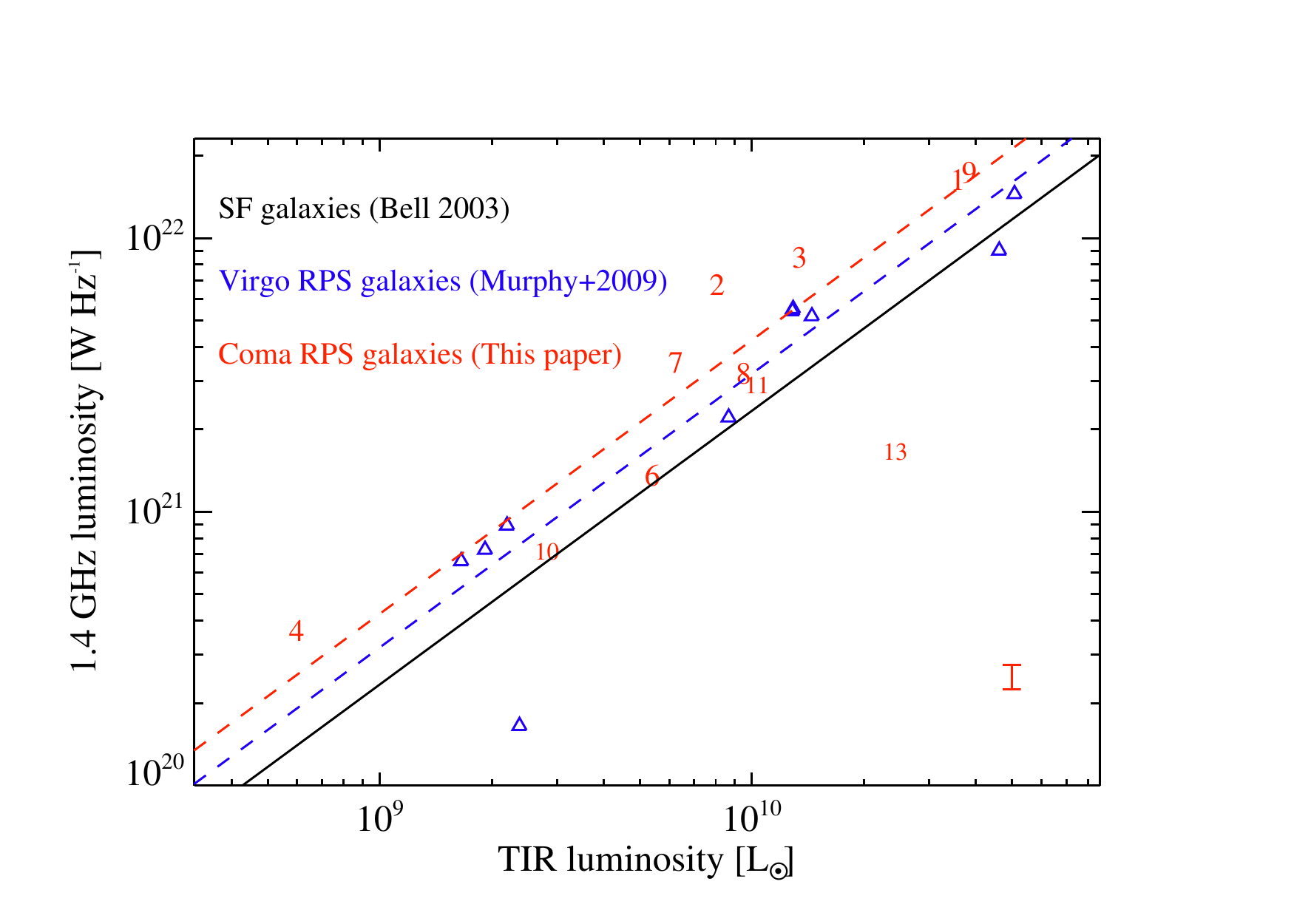}}
	\vspace{-1cm}
	\centerline{\includegraphics[width=1.13\columnwidth]{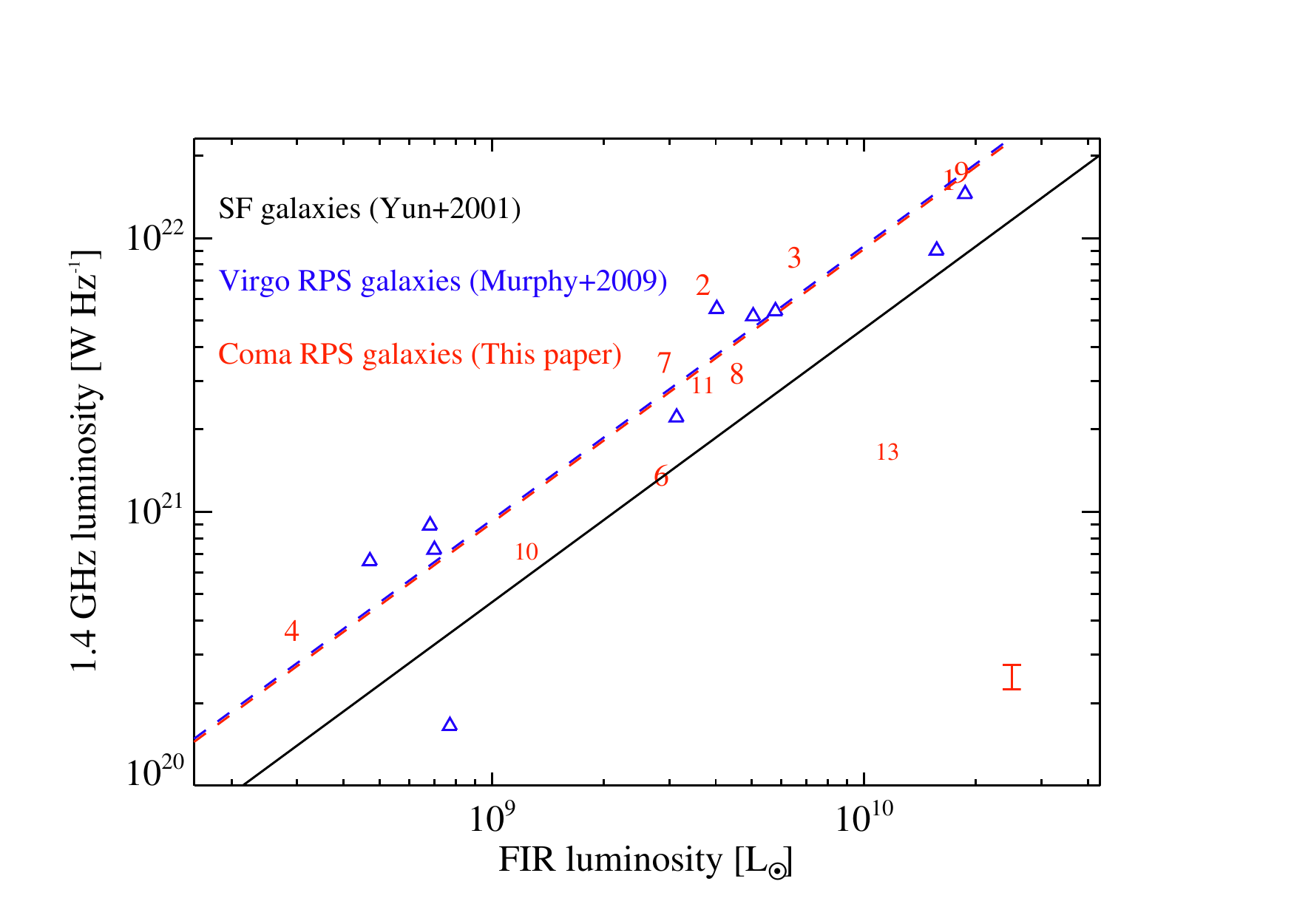}}
	\vspace{-0.3cm}
    \caption{The 1.4 GHz luminosity versus total infrared (TIR, top plot) and far-infrared (FIR, bottom plot) luminosity for the RPS galaxies. The galaxies in the Coma cluster are in red and those in the Virgo cluster \citep{Murphy2009ApJ...694.1435M} are in blue. Each number indicates a galaxy as listed in Table~\ref{tab:tail_sources_table}. The mean ratios of 1.4 GHz to TIR (FIR) for RPS galaxies in the Coma cluster and in the Virgo cluster are shown as red and blue dashed lines respectively. The relations between the 1.4 GHz continuum and TIR \citep[][]{Bell2003ApJ...586..794B} or FIR \citep[][]{Yun2001ApJ...554..803Y} for star forming galaxies are also shown as black solid lines.}
    \label{fig:cont-IR}
\end{figure}

There is also evidence that the ram pressure may also enhance the radio emission of the main bodies of the galaxies. With the 70, 100 and 160 $\micron$ flux densities for these RPS galaxies from the {\em Herschel} point source catalog, the radio-IR relations of these RPS galaxies are compared with those for star forming galaxies (Fig.~\ref{fig:cont-IR}). Because the main bodies are much brighter than the tails, the radio and IR emissions of the RPS galaxies are dominated by their main bodies.
The TIR luminosity of the RPS galaxies in the Coma cluster is derived from the 70, 100 and 160 $\micron$ flux densities with the calibration coefficients listed in Table 3 of \citet{Galametz2013MNRAS.431.1956G}. The corresponding far-infrared (FIR, $42-122\ \micron$) fraction was given by the 70 to 160 $\micron$ flux density ratio with the Equation 3 of \citet{Murphy2008ApJ...678..828M}.
The IR to 1.4 GHz ratio is generally expressed by the parameter $q$ \citep[][]{Bell2003ApJ...586..794B,Helou1985ApJ...298L...7H}:
\begin{equation}
{\rm {\textit q}_{IR} = log_{10}(\frac{IR}{3.75\times10^{12}\ W\ m^{-2}})-log_{10}(\frac{{\textit S}_{1.4\ GHz}}{W\ m^{-2}\ Hz^{-1}})},
\end{equation}
The mean values of $q_{\rm TIR}$ and $q_{\rm FIR}$ in the RPS galaxies of the Coma cluster ($q_{\rm TIR}=2.38 \pm 0.25$ and $q_{\rm FIR}=2.05 \pm 0.24$, the red dashed lines in Fig.~\ref{fig:cont-IR}) are systematically lower than that in star forming galaxies \citep[$q_{\rm TIR}=2.64 \pm 0.26$ and $q_{\rm FIR}=2.34 \pm 0.26$, the black solid lines in Fig.~\ref{fig:cont-IR},][]{Bell2003ApJ...586..794B,Yun2001ApJ...554..803Y}.  
On average, the 1.4 GHz continuum in the RPS galaxies in the Coma cluster is enhanced relative to their TIR (by a factor of 1.9) and FIR (by a factor of 2.0) compared to normal star-forming galaxies. 
Previous works also reported the radio excess in the Coma cluster galaxies \citep[e.g.,][]{Gavazzi1986ApJ...310...53G,Gavazzi1991AJ....101.1207G} and in the Virgo cluster galaxies \citep[e.g.,][]{Murphy2009ApJ...694.1435M,Vollmer2013A&A...553A.116V}.
The mean $q_{\rm TIR}$ and $q_{\rm FIR}$ in the Virgo cluster galaxies are $2.50 \pm$ 0.12 and $2.04 \pm 0.20$ \citep[the blue dashed lines in Fig.~\ref{fig:cont-IR}, ][]{Murphy2009ApJ...694.1435M}. 
Several mechanisms have been proposed to explain the enhanced radio continuum in RPS galaxies, including the enhancement of magnetic field strength caused by ISM shear motions \citep{Murphy2009ApJ...694.1435M}, the compression of ISM/magnetic field by the ICM ram and/or thermal pressure \citep{Murphy2009ApJ...694.1435M}, and turbulence and ICM shocks caused by ram pressure \citep{Kang2011ApJ...734...18K}.
The general radio enhancement in the Coma tails may be related to turbulence in the tail and wake of the galaxy, a further investigation of which would require more theoretical or simulation work which is beyond the scope of the current paper.

\subsubsection{Synchrotron cooling time}

If the 1.4 GHz emission in the stripped tail originates from the relativistic electrons stripped from the galaxy by ram pressure, the relativistic electrons will cool/age along the tail. The age of the relativistic electrons in the tail, $t$, is roughly $d/v$, where $d$ is the distance from the galaxy and $v$ is the mean velocity of electrons stripped from the galaxy. The 1.4 GHz emission can only be detected before the relativistic electrons have cooled, or roughly $t < t_{\rm syn}$(1.4\ GHz), which is equivalent to $d < t_{\rm syn}{\rm (1.4\ GHz)}\times v$, where $t_{\rm syn}$(1.4\ GHz) is the synchrotron cooling time at 1.4 GHz. With this assumption, if the tail length detected at 1.4 GHz does not exceed $t_{\rm syn}{\rm (1.4\ GHz)}\times v$, it will support the assumption that the bulk of the radio continuum in tails is stripped from the galaxy.

Using equation 9 of \citet{Feretti2008LNP...740..143F}, the synchrotron cooling time (in Myr) can be estimated as:
\begin{equation}
\label{tausyn2} 
t_{\rm syn} (\nu) = 1590 \frac{B^{0.5}}{B^2+B_{\rm CMB}^2} [(1+z)\nu]^{-0.5} 
\end{equation}
where the magnetic field $B$ is in $\mu$G, the frequency is in GHz and ${B_{\rm CMB}}=3.25(1 + z)^2$ $\mu$G is the magnetic field of the Cosmic Microwave Background.
Faraday rotation measure studies indicate that magnetic field
strengths in clusters are on the order of a few $\mu$G, with strengths
up to tens of $\mu$G in cluster cores \citep[e.g.,][]{Perley1991AJ....101.1623P,Taylor1993ApJ...416..554T,Feretti1995A&A...302..680F,Feretti1999A&A...344..472F,Taylor2002MNRAS.334..769T,Taylor2006MNRAS.368.1500T,Taylor2007MNRAS.382...67T,Bonafede2010A&A...513A..30B}. However, if the magnetic field in the tail has its origin in the galaxy, it would be stronger than the magnetic field in the ICM.
Assuming a magnetic field strength of 10 $\mu$G, $t_{\rm syn}$ will be $\sim38$ Myr at 1.4 GHz for the Coma cluster (z=0.0231).  
The mean velocity of the relativistic electrons stripped from the galaxy is about 500 km\,s$^{-1}$ in simulations \citep{Tonnesen2010ApJ...709.1203T}. 
Assuming an isotropic distribution of tail directions, the velocity in the plane of the sky is $\sim$400 $\rm{km\ s^{-1}}$ ($500\times\pi/4\ \rm{km\ s^{-1}}$).
Then, the maximum expected length of the radio continuum tail at 1.4\,GHz in our observations would be 15.2 kpc, which is
indeed larger than the observed length of the radio continuum tails (2.8--12.6 kpc) in the Coma cluster.
The observed steepening of radio spectra in tails (especially for NGC~4858) also supports synchrotron cooling in the tail.
The large uncertainty in the assumed relative velocity of the galaxy with respect to the ICM and the value for the magnetic field lead to a large range for the maximum length limitation of the radio continuum tails and the above should thus be considered to be an order of magnitude calculation based on plausible values.

\subsubsection{RPS galaxies without radio tails}

There are 10 galaxies without radio continuum tail detections. For 5 of them (GMP~4060, NGC~4854, GMP~3016, GMP~2923 and KUG~1258+277), no radio continuum is detected at all from the galaxy (Table~\ref{tab:tail_sources_table}). 
There is no deep H$\alpha$ data for KUG~1258+277.
The H$\alpha$ tails of the other 4 \citep[GMP~4060, NGC~4854, GMP~3016 and GMP~2923, corresponding to Fig.~4 l, k, e and d in][]{Yagi2010AJ....140.1814Y} are weak and detached from the galaxies.
At the same time, GMP~4060, GMP~3016, and GMP~2923 did not show detectable H$\alpha$ emission in the galaxies.
So perhaps they are at a late evolutionary stage of stripping which might account for the weak (undetected) radio emission.

There are 5 galaxies (NGC~4911, NGC~4921, NGC~4853, IC~3949 and GMP~3071) with radio continuum detections in the galaxies, but radio continuum tails are not detected. 
NGC~4911 and NGC~4921 are both face-on galaxies. 
Their H$\alpha$ extensions are short, which is consistent with the lack of long radio tails there.
Whereas NGC~4911 does not have a significant radio tail, its radio continuum within $D_{25}$ is asymmetric, with an extension towards the short H$\alpha$ tail. NGC~4853, IC~3949 and GMP 3071 do have significant H$\alpha$ tails, but no radio continuum tails are detected. Deeper data may be required to probe the radio continuum emission in the tails of these galaxies.

\subsection{HI deficiency}

\subsubsection{HI deficiency of RPS galaxies}

The RPS galaxies should be deficient in HI as it is stripped \citep{Quilis2000Sci...288.1617Q,Boselli2016A&A...587A..68B}.
To quantify the missing atomic gas in the RPS galaxies in Coma, the HI deficiency parameter \citep[Def$_{\rm HI}$,][]{Haynes1984AJ.....89..758H}, or the log of the ratio of the normal (i.e., expected) HI mass and the observed HI mass, is studied with our HI results. 
The whole HI flux within $D_{25}$ is used to derive the Def$_{\rm HI}$ for NGC~4848, NGC 4911 and IC 4040.
For other galaxies, only a lower limit within $D_{25}$ can be placed on Def$_{\rm HI}$.
The upper limit on the HI flux is
\begin{equation}
S_{\rm HI\_ul} = 3 \times rms \times \sqrt{W\times W_0} \times \sqrt{A_{\rm galaxy}/A_{\rm beam}}
\ ,
\end{equation}
where rms is the local noise in the HI data, in units of Jy beam$^{-1}$ channel$^{-1}$, $W$ is the expected velocity width of the galaxy in units of ${\rm km\ s^{-1}}$, $W_0$ is the channel width of our data cube (21.5 ${\rm km\ s^{-1}}$), $A_{\rm galaxy}$ is the $D_{25}$ area of the galaxy, and $A_{\rm beam}$ is the area of the beam size for our HI data cube.
The expected velocity width of the galaxy is given by $W=2(V_{\rm M}\ {\rm sin}~i)+W_{\rm t}$ \citep{Bottinelli1983A&A...118....4B}, which is derived from the maximum rotational velocity ($V_{\rm M}$) by correcting the projection of inclination ($i$) and profile broadening ($W_{\rm t}$) caused by random/turbulent motions. $V_{\rm M}$ is obtained from $m_{\rm B}$ using the Tully-Fisher relation for individual galaxies \citep[Fig.~13 of][]{Meyer2016MNRAS.455.3136M}. The values of $m_{\rm B}$ and $i$ are from LEDA and listed in Table~\ref{tab:HI_Deficiency}. The value of $W_{\rm t}$ used here is 5 ${\rm km\ s^{-1}}$ from \citet{Verheijen2001A&A...370..765V}.
The HI mass is given by \citet{Meyer2017PASA...34...52M}:
\begin{equation}
M_{\rm HI} =  \frac{2.35 \times 10^5}{(1+z)^2} \times D^2 \times S_{\rm HI}
,
\end{equation}
where $M_{\rm HI}$ is the HI mass in M$_\odot$, $S_{\rm HI}$ is the HI flux in ${\rm Jy\ km\ s^{-1}}$, $D$ is the luminosity distance of the galaxy in Mpc (100.7 Mpc used for all the Coma galaxies), and $z$ is the redshift of the galaxy (0.0231 used for all the Coma galaxies). 
The normal or expected HI mass ($M_{\rm HI}$$_{\rm normal}$) is derived based on table IV of \citet{Haynes1984AJ.....89..758H}, on $D_{25}$ (LEDA data base) and the galaxy type \citep[][LEDA data base]{Dressler1980ApJS...42..565D}. 
All the results are listed in Table~\ref{tab:HI_Deficiency}.

\begin{table*}
    \caption{HI in the stripped tail of D100}
	\begin{tabular}{ccccccccc}      
		\hline
		 Tail Region &  Area\_HI & FWHM$_{\rm CO}$ & $M_{\rm HI}$  & $M_{\rm H_2}$ & $R_{\rm H2}$ & P/k$_{\rm B}$ \\
                          &  & [km s$^{-1}$] & [$10^8 {\rm M_{\odot}}$]& [$10^8 {\rm M_{\odot}}$] &    & [10$^5$ K cm$^{-3}$] \\
		\hline
        	D100\_T1 	 &$4\arcsec\times10.8\arcsec$ &145.5 &$<0.95$ & $0.73\pm0.09$ &$>$ 0.77 & 1.9  &  &\\
        	D100\_T2 	 &$4\arcsec\times21.7\arcsec$ &165.3 &$<$ 1.43 & 3.01 $\pm$ 0.44 &$>$ 2.11 & 1.9  &  &\\
        	D100\_T3 	 &$4\arcsec\times21.7\arcsec$ &52.8  &$<$ 0.81 & 1.97 $\pm$ 0.22 &$>$ 2.45 & 1.9  &  &\\
        	D100\_T4 	 &$4\arcsec\times21.7\arcsec$ &31.3  &$<$ 0.62 & 0.28 $\pm$ 0.16 &$>$ 0.45 & 1.9  &  &\\
            D100\_Tail 	 &$4\arcsec\times108.5\arcsec$&200.0 &$<$ 3.51 & $>$ 5.98        &$>$ 1.71 & 1.9  &  &\\
        \hline
	\end{tabular}
    \begin{tablenotes}
    \item {\sl Note:} The HI mass upper limit in the stripped tail of D100 is derived from the data presented here. The size of the HI tail is set by the H$\alpha$ emission ($4\arcsec\times108.5\arcsec$);
the velocity width is assumed to be the same as CO (FWHM$_{\rm CO}$).
D100 $M_{\rm H2}$ results adjusted to the cosmology of this paper are from CO 1--0 \citep{Jachym2017ApJ...839..114J}. The exception is the tail region of T1 where we used CO 2--1 because the beam of CO 1--0 covered part of disk.
P/K$_{\rm B}$ is the ICM thermal pressure at the projected position of D100.
    \end{tablenotes}
    \label{tab:D100_H2/HI}
\end{table*}

As the HI disks of galaxies within $\sim$ 1 Mpc of the centre of clusters are typically truncated to well within the $D_{25}$ \citep[e.g.,][]{Chung2009AJ....138.1741C}, our HI deficiencies should therefore be accurate.
Def$_{\rm HI}$ for IC 4040 and the lower limits on Def$_{\rm HI}$ for NGC 4858, MRK 0058, NGC 4921, KUG 1258+279A and KUG 1258+277 are consistent with \citet{Bravo-Alfaro2000AJ....119..580B} (Def$_{\rm HI\_BA00}$).
However, in this work we found NGC~4848 and NGC~4911 to be less HI-deficient than \citet{Bravo-Alfaro2000AJ....119..580B}.
This is because the HI flux is larger than that in \citet{Bravo-Alfaro2000AJ....119..580B} which is also discussed in section 3.3.
Most (15 of 20) RPS galaxies in our sample are proved to be HI deficient with our detection and lower limit estimate. 
Having said that, we should keep in mind that $M_{\rm HI}$ could be an underestimate as any missing flux has not been considered here.

\subsubsection{HI in the stripped tail of D100}

The mixing of the stripped cold ISM with the hot ICM will produce multi-phase gas \citep[e.g.,][]{Sun2007ApJ...671..190S,Ferland2009MNRAS.392.1475F,Jachym2019ApJ...883..145J}.
It is interesting to study how multi-phase gases co-exist and evolve in the RPS tails.
D100 is the only galaxy in our sample with CO detections in the RPS tails \citep{Jachym2017ApJ...839..114J}.
With our deep and high spatial resolved HI observations, the state of co-exist molecular and atomic gas in the RPS tails of D100 could be discussed.

We were able to derive the molecular gas mass, $M_{\rm H_2}$, in the narrow tail \citep[regions D100\_T1 to D100\_T4 in][]{Jachym2017ApJ...839..114J} of D100 from CO single dish observations and Galactic standard CO--${\rm H_2}$ relation.
The lower limit of $M_{\rm H_2}$ for the whole tail could be got by the sum of D100\_T1 to D100\_T4 as the tail is not fully covered by the CO observations.
We constrain the HI mass limit in the narrow H$\alpha$ tail of D100 ($\sim 4\arcsec$ width) with the new data ($\sim 6 \arcsec$ spatial resolution vs. $\sim 30 \arcsec$ spatial resolution for \citet{Bravo-Alfaro2000AJ....119..580B}).
The HI mass upper limit is derived using equation 2 and 3 for the whole tail (4\arcsec$\times$108.50\arcsec) assuming that the HI velocity width is the same as CO (the velocity coverage of CO in D100\_T1 to D100\_T4 is about 200 ${\rm km\,s^{-1}}$).
The HI upper limit is also derived for each part of the H$\alpha$ tail covered by CO observations (D100\_T1 to D100\_T4).
Results are listed in Table~\ref{tab:D100_H2/HI}.
Integrated over the entire tail there is more than $5.98\times10^8$ M$_\odot$ of H$_2$ (assuming a Galactic CO-to-H2 conversion factor), and less than $3.51\times10^8$ M$_\odot$ of HI. 
The molecular gas fraction in (parts of) D100's stripped tail is surprisingly high, which is normally only observed in the inner disk of galaxies. This is consistent with the results in (parts of) the RPS tail of ESO137-001 \citep{Jachym2014ApJ...792...11J}.

In galaxies, the molecular to atomic gas ratio ($R_{\rm H_2}$) correlates with the interstellar pressure in the galactic disk \citep[][]{Blitz2004ApJ...612L..29B,Blitz2006ApJ...650..933B}.
The ICM thermal pressure (${\rm P/k_B}$) at the projected position of D100 (0.17 Mpc away from the cluster centroid) is ${\rm 1.9\times 10^5\ K\ cm^{-3}}$, which is derived from model C of the \citet{Planck2013A&A...554A.140P}.
Although the environments are completely different, the high molecular to atomic gas ratio ($R_{\rm H_2}$) and high ICM thermal pressure in (parts of) D100's stripped tail agree well with the typical relation between $R_{\rm H_2}$ and pressure in the galactic disk \citep[Fig. 9 in][]{Krumholz2009ApJ...693..216K}.
The high ICM thermal pressure offers an explanation for the molecular gas dominated stripped tail \citep{Jachym2014ApJ...792...11J}. On the other hand, formation of molecular gas and ionization of atomic gas in the stripped tail also increase $R_{\rm H_2}$.

\section{Conclusions}

We obtained deep VLA 1.4\,GHz data in two fields of the Coma cluster and focused on 20 RPS galaxies by studying their radio continuum and HI properties. Our main results are summarized as follows: 

(1) Radio continuum tails are found in 10 of 20 ram pressure stripped (RPS) galaxies in Coma, revealing the widespread occurrence of relativistic electrons and magnetic fields in the RPS tails. 8 of the RPS tails are new detections.

(2) The wide-band 1.4\,GHz data allow spectral indices to be derived which provide a hint that the radio spectra of the tails are steeper than those of the galaxies, as expected from synchrotron cooling without injection of fresh relativistic electrons in the tails.

(3) The 1.4\,GHz continuum of the stripped tails is enhanced relative to their SFR traced by H$\alpha$ emission by a factor of $\sim$ 7 compared to the main bodies of the galaxies.
It suggests that the radio continuum in the stripped tail is not related to the local SF activity, but rather related to the ram pressure stripping.

(4) Radio emission of the RPS galaxies in the Coma cluster is also enhanced relative to its TIR (or FIR) emission by a factor of $\sim$ 2, compared with normal SF galaxies. Ram pressure interaction may enhance radio emission in the main bodies of the galaxies.

(5) The length of the radio continuum tails is consistent with the distance being limited by the synchrotron cooling time and stripping velocity.

(6) HI detections in three RPS galaxies (NGC~4848, NGC~4911 and IC~4040) and HI upper limits for the other galaxies are presented as well. Most (15 of 20) RPS galaxies are consistent with being deficient in HI.

(7) The H$_2$/HI mass ratio for the H$\alpha$ tail of D100 exceeds 1.6 --- the cold gas in D100's stripped tail is dominated by molecular gas. This is likely a consequence of the high ICM pressure in the stripped tail.

Our continuum data imply that there are magnetic fields in the stripped tails, which could explain the collimated and bifurcated structure of the stripped gas.
SF may be less efficient in a stripped tail due to suppression by magnetic fields and turbulence. 
We suggest that the observed radio continuum emission is due to ram pressure stripping of relativistic electrons from the host galaxy, and not so much to any local SF. This is consistent with our finding of a steeper radio spectrum in the stripped tail compared to that of the host galaxy, the overly luminous radio continuum flux density compared to the H$\alpha$, and the shorter tail length in radio (limited by the synchrotron cooling time) than in H$\alpha$. Our results also suggest that cluster late-type galaxies can inject relativistic electrons into the ICM, which could potentially feed radio relics in clusters \citep{Ge2019MNRAS.486L..36G}.

\section*{Acknowledgements}

We acknowledge useful discussions with the NRAO help desk staff, Xing Lu, Ruoyu Liu, Neal Miller and Stephen Walker. Support for this work was provided by NSF grant 1714764 and  NASA/EPSCoR grant NNX15AK29A. We also acknowledge the support from the National Aeronautics and Space Administration through {\em Chandra} Award Number GO6-17111X and GO6-17127X issued by the {\em Chandra} X-ray Center, which is operated by the Smithsonian Astrophysical Observatory for and on behalf of the National Aeronautics Space Administration under contract NAS8-03060. 
HC's work has been supported by the South African Department of Science and Innovation and the National Research Foundation through a Fellowship within the SARAO Research Chair held by RC Kraan-Korteweg.
MF has received funding from the European Research Council (ERC) under the European Union's Horizon 2020 research and innovation programme (grant agreement No 757535).
This research has made use of data and/or software provided by the High Energy Astrophysics Science Archive Research
Center (HEASARC), which is a service of the Astrophysics Science Division at NASA/GSFC and the High Energy Astrophysics Division
of the Smithsonian Astrophysical Observatory. This research has made use of the NASA/IPAC Infrared Science Archive, which is funded by the National Aeronautics and Space Administration and operated by the California Institute of Technology.
Based in part on data collected at Subaru Telescope, which is operated by the National Astronomical Observatory of Japan.

\section*{Data availability}

Most of the data underlying this article are available in the article and in its online supplementary material. All the data will be shared on reasonable request to the corresponding authors.




\bibliographystyle{mnras}
\bibliography{coma_radio} 




\appendix

\section{Radio continuum map}

\begin{figure*}
\begin{minipage}{0.63\linewidth}
\includegraphics[width=1.03\textwidth]{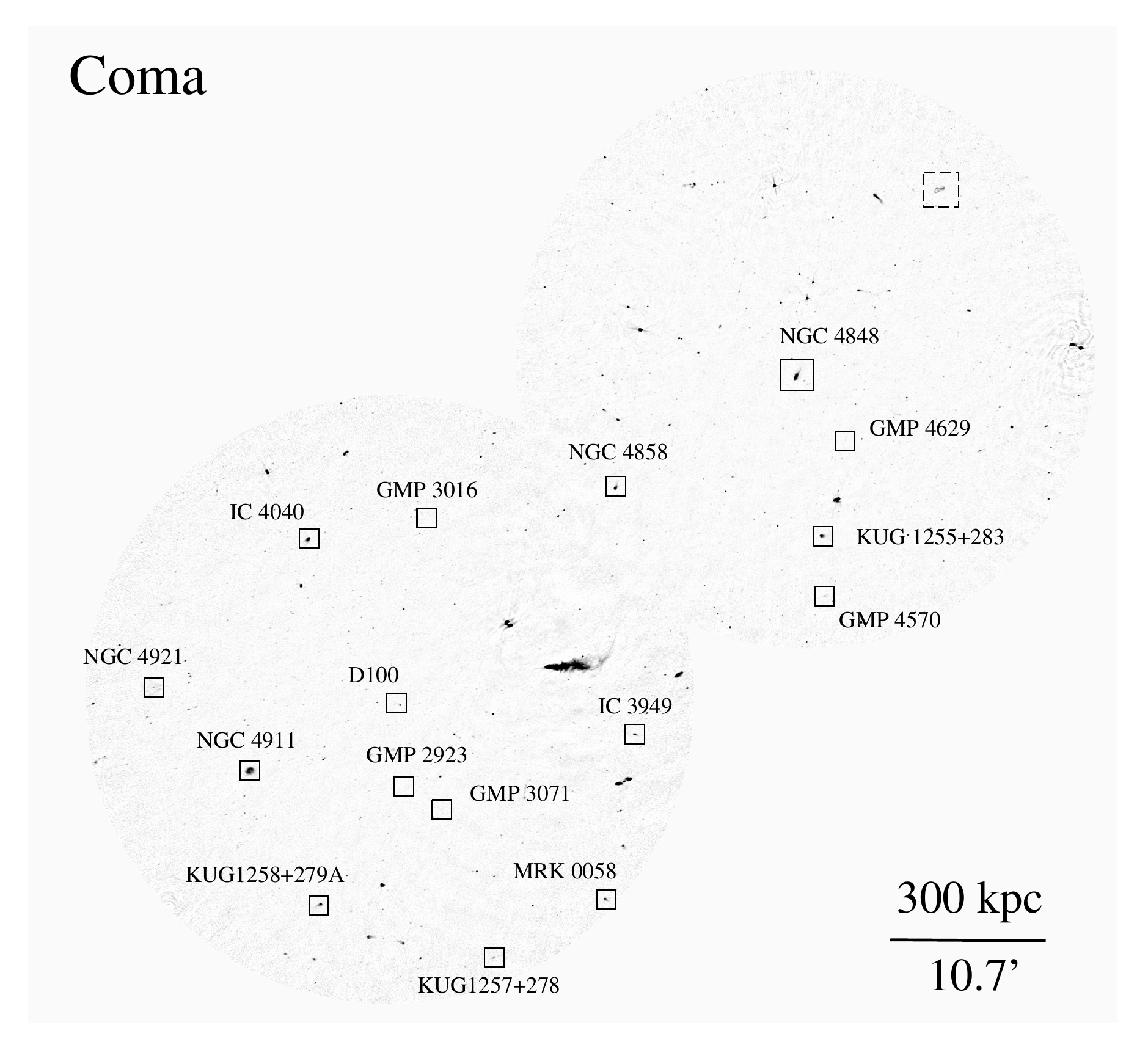} 
\end{minipage}
\hfill
\begin{minipage}{0.35\linewidth}
  \includegraphics[width=0.97\textwidth]{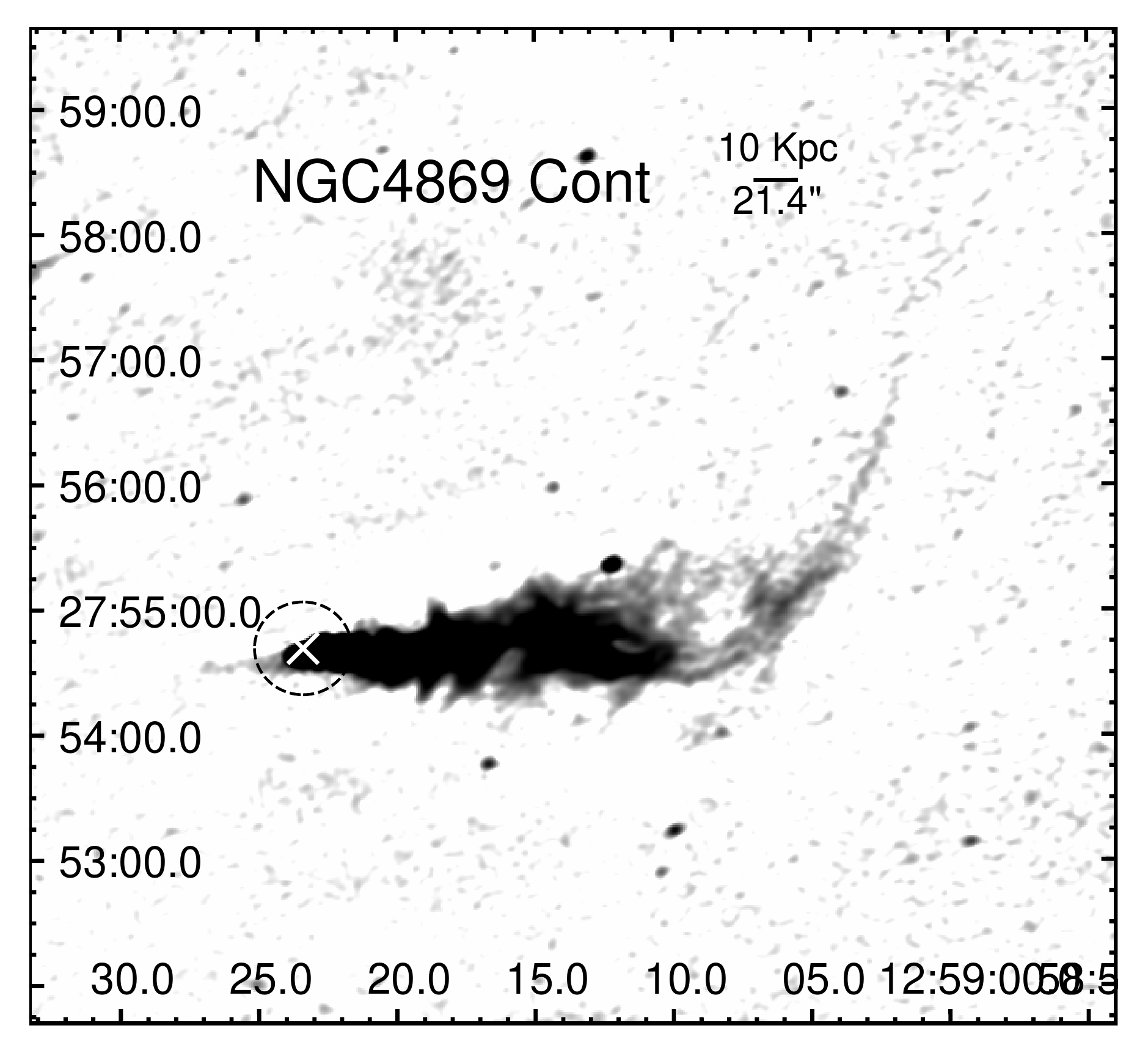}
  \includegraphics[width=0.97\textwidth]{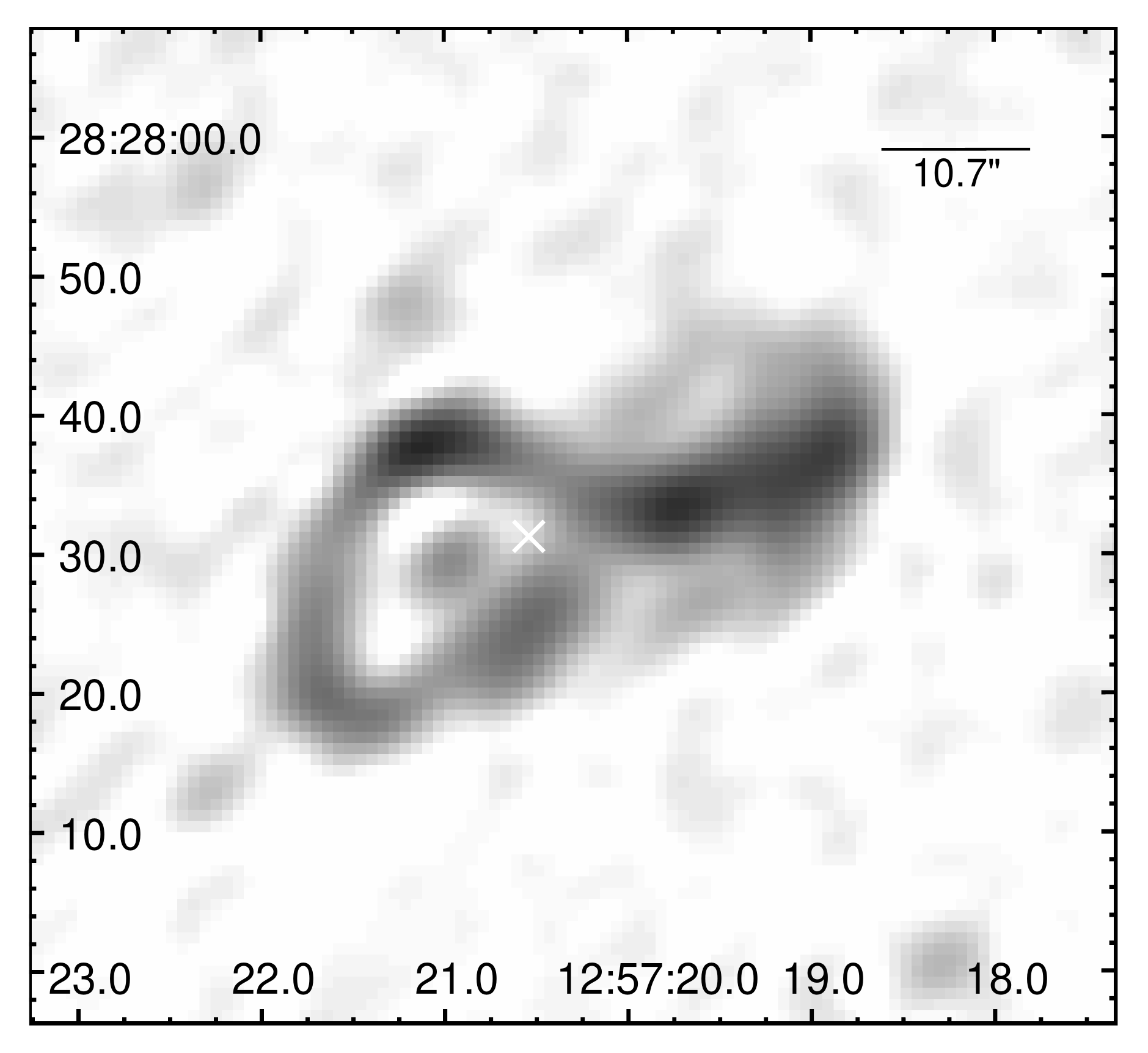}
\end{minipage}
	 \caption{Left: Radio continuum image of two fields in the Coma cluster (30\% response shown here). Galaxies shown in Fig.~\ref{figure:coma} and also covered by 30\% response are marked as solid squares. Right: zoom-in radio continuum images of two interesting sources: the narrow-angle tailed galaxy NGC~4869, and an interesting source at 12:57:20.54 +28:27:31.28 (J2000.0) marked as a dashed square.
	 }
	\label{fig:radio_map}
\end{figure*}

A mosaic image of our radio continuum observations is shown in Fig.~\ref{fig:radio_map}. It is the combination of the 30\% response regions of the two fields.
Magnified images of two sources are shown at the right to show them in detail.
The narrow tail near the end of NGC~4869 is shown better in our image than in \citet{Miller2009AJ....137.4436M}.
The source in the lower right panel (the small dashed-line box on the left) has an interesting complex structure which is not covered by \citet{Miller2009AJ....137.4436M}.
This source is cataloged as 1254+28W07 \citep{Valentijn1977A&AS...28..333V,Kim1994A&AS..105..403K} and NVSS J125720+282727 \citep{Condon1998AJ....115.1693C}, but the complex structure is not resolved because of the low spatial resolution ($>45 \arcsec$).

\section{Flux density measurements}

\begin{table*}
	\caption{The full continuum source catalog}
	\begin{tabular}{ccccccccc} 
		\hline
		 ID (1) & R.A. (J2000) (2) & Decl. (J2000) (3) & $S_{\rm int}$ (4) & $\Delta S_{\rm int}$ (5) & field (6) \\
                      & [h m s] & [$^\circ$\ $\arcmin$\ $\arcsec$] & ${\rm [\mu Jy]}$ & ${\rm [\mu Jy]}$ & \\
		\hline
         1&      12      56      2.35&      28      15     28.6&        181&       60&      1\\
         2&      12      56      2.73&      28      13     15.6&        141&       57&      1\\
         3&      12      56      2.77&      28      15     24.6&        169&       93&      1\\
         4&      12      56      3.68&      28      15     49.9&        217&       72&      1\\
         5&      12      56      3.69&      28      20     52.4&        123&       47&      1\\
\hline
	\end{tabular}
    \begin{tablenotes}
    \item {\sl Note:}
Column descriptions: (1) source ID; (2) and (3) right ascension and declination (J2000) of radio source peak; (4) and (5) integral flux density and its associated error at 1.4 GHz; (6) The field (shown in Fig.~\ref{figure:coma} and Table~\ref{tab:observation_table}) in which source was found. For the 64 sources covered by both fields, their results from the two fields are listed in different rows respectively and the same source ID was set for the same source. 
\\
\ \\
(This table is available in its entirety in machine-readable form online. A portion is shown here for guidance regarding its form and content.)
    \end{tablenotes}
     \label{tab:catolog_table}
\end{table*}

\begin{figure*}
	\includegraphics[width=2\columnwidth]{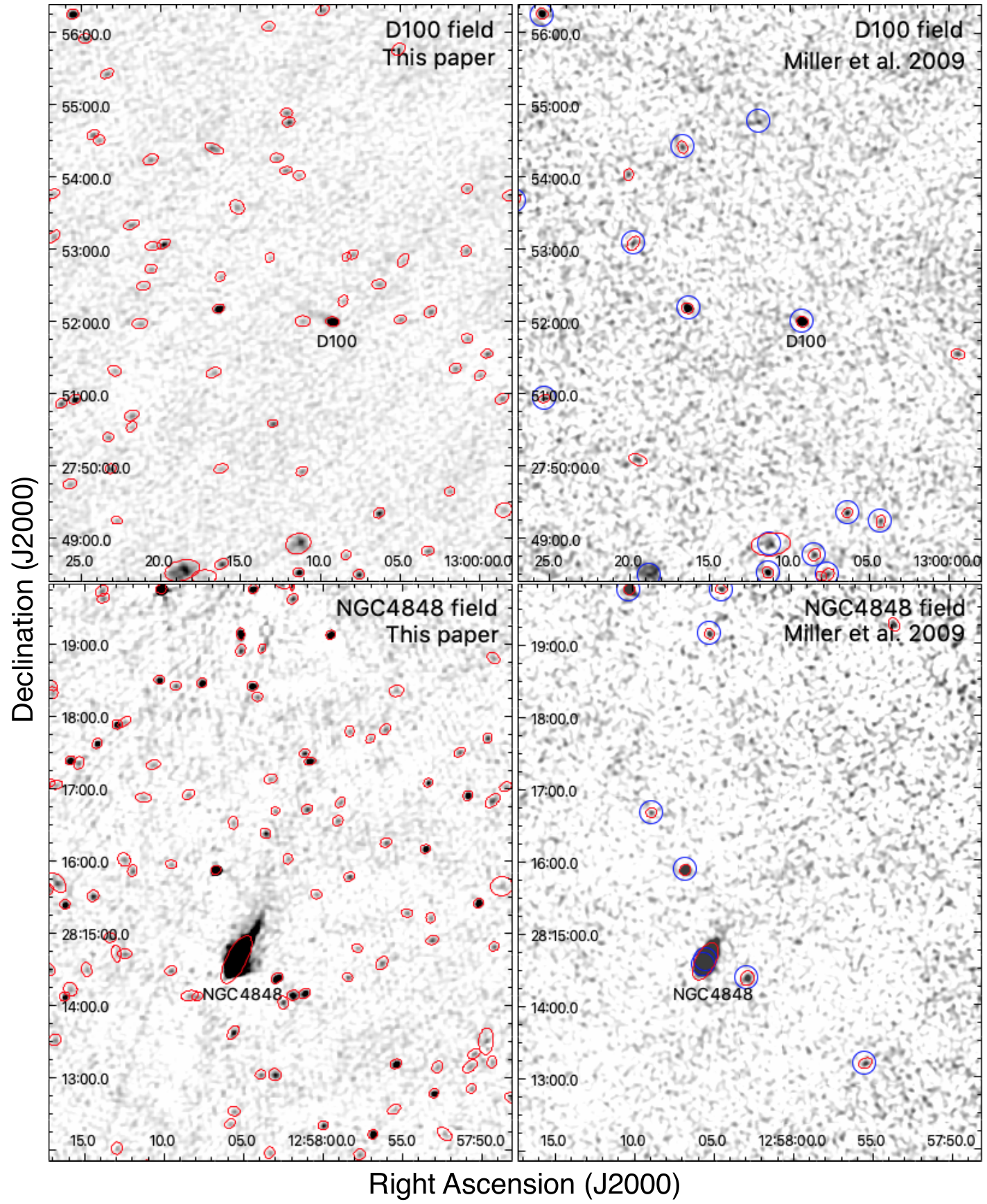}
	\vspace{-0.3cm}
    \caption{Comparison of sources detected with PyBDSF between this work and \citet{Miller2009AJ....137.4436M}. Red apertures show the radio continuum sources defined with PyBDSF. Blue circles show the sources detected by \citet{Miller2009AJ....137.4436M} (the radii are fixed at 20$\arcsec$).
    }
    \label{fig:compare cont image}
\end{figure*}

\begin{figure}
	\includegraphics[width=1\columnwidth]{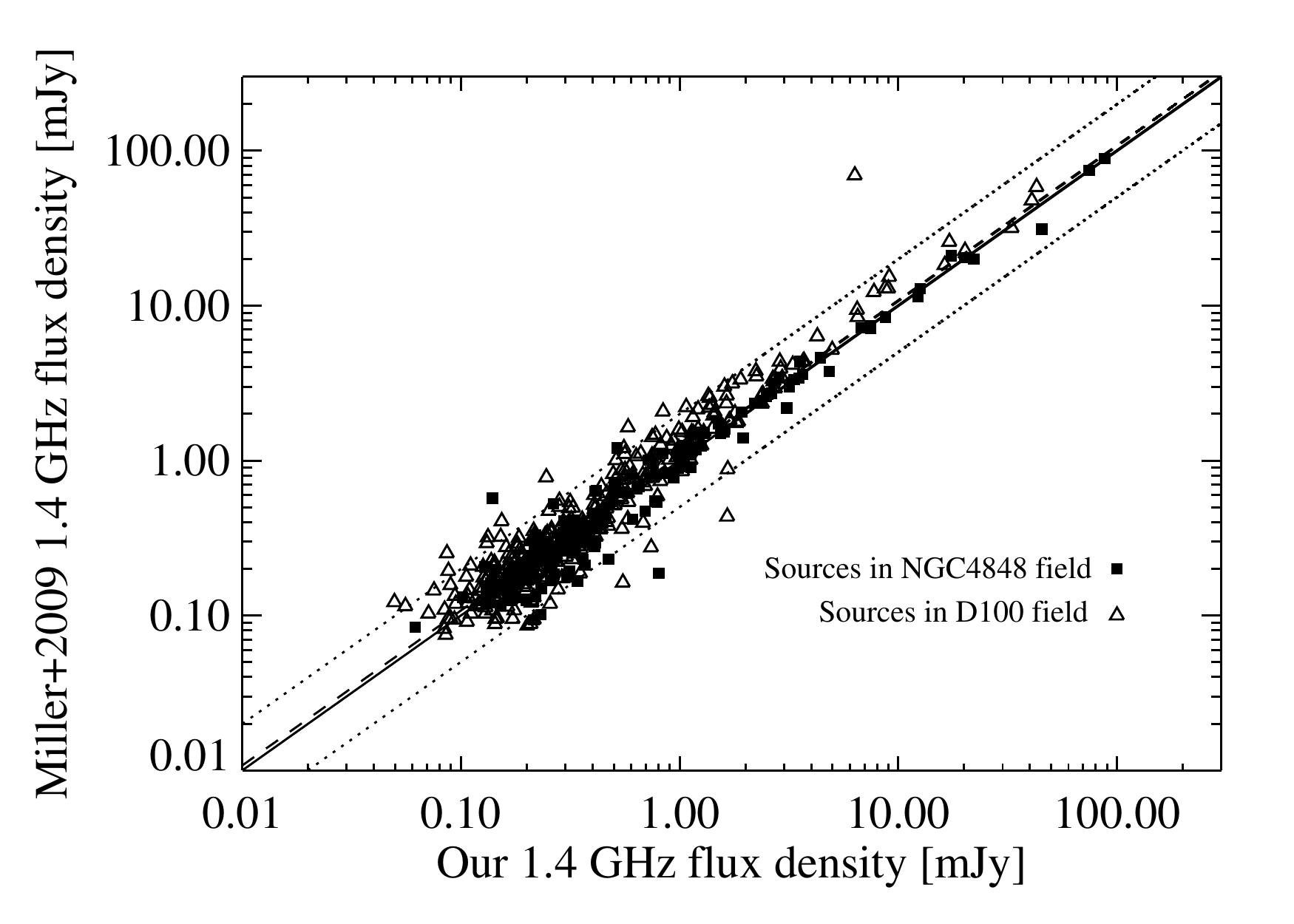}
	\includegraphics[width=1\columnwidth]{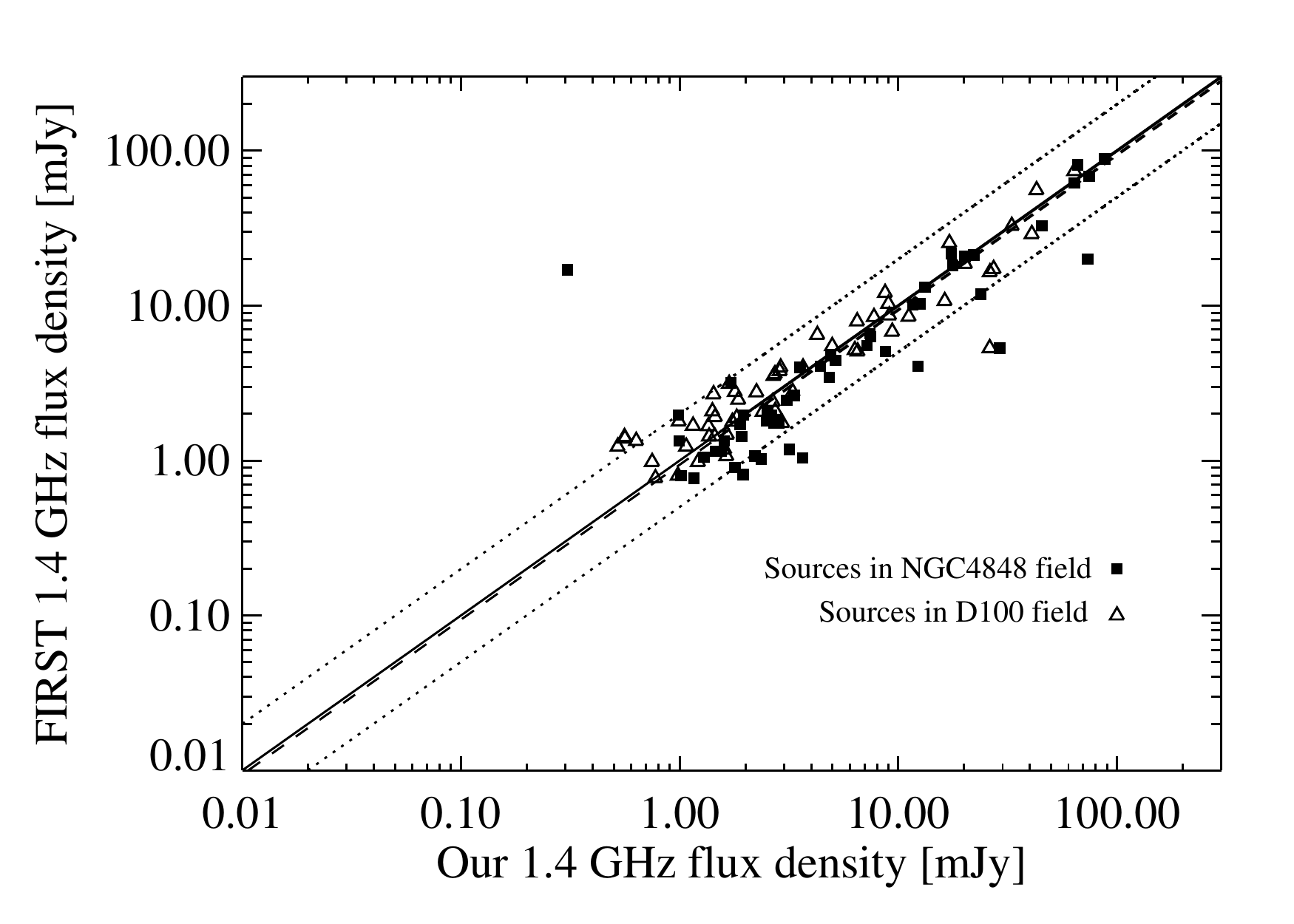}
    \includegraphics[width=1\columnwidth]{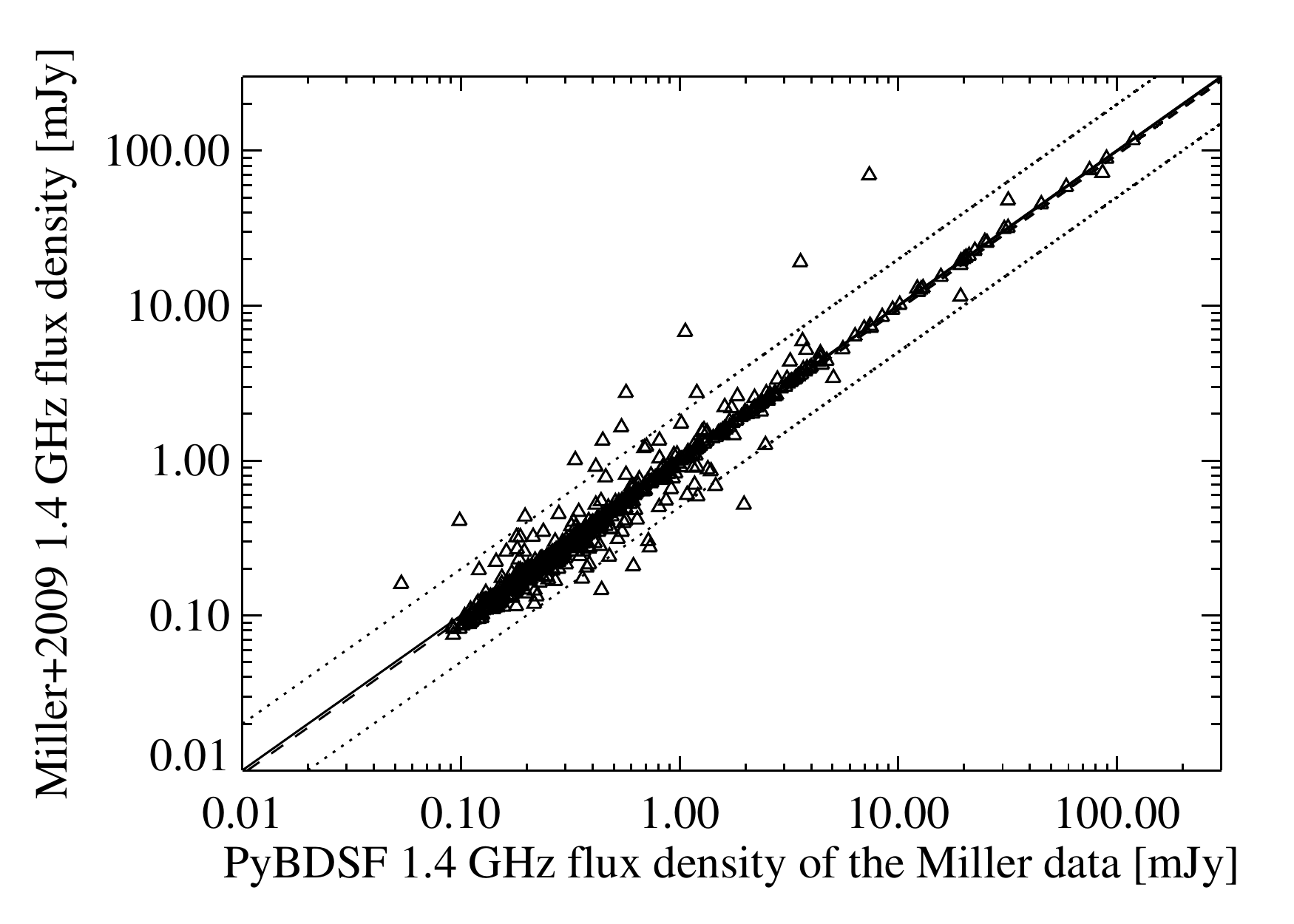}
    \caption{Top two plots show the flux density comparison between this work and \citet{Miller2009AJ....137.4436M} and the FIRST survey \citep{White1997ApJ...475..479W}.
    The bottom plot shows the comparison between \citet{Miller2009AJ....137.4436M} and PyBDSF results on their data.
    Dotted lines show an uncertainty of 0.3 index,
    while dashed and solid lines show the linear and power-law fitting results.
    }
    \label{fig:compare flux}
\end{figure}

The full continuum source catalog is shown in Table~\ref{tab:catolog_table}.
As Fig.~\ref{fig:compare cont image} shows, our deep observations reveal many more continuum sources than \citet{Miller2009AJ....137.4436M}. 
To assess the reliability of the flux density measurement of our work, we compared our results with 
\citet[][]{Miller2009AJ....137.4436M} and the FIRST survey \citep{White1997ApJ...475..479W}. 
Because the spatial resolution is about 5$\arcsec$, we cross match Miller's and FIRST sources with ours for match radii of 5$\arcsec$.
Sources are excluded when more than one match is found within 5$\arcsec$. The comparisons are shown in Fig.~\ref{fig:compare flux}.
Our flux density measurements are consistent with those from \citet{Miller2009AJ....137.4436M} and FIRST.
There is still some scatter, which could be caused by: 1) different algorithms of source detection (PyBDSF in our work and SAD in \citet{Miller2009AJ....137.4436M} and FIRST); and/or 2) varying radio sources (e.g., AGN). 
We also applied the PyBDSF algorithms to the data of \citet{Miller2009AJ....137.4436M} to compare the flux density measurements. 
By applying PyBDSF on both our data and the \citet{Miller2009AJ....137.4436M} data, the consistency on the flux density measurements is improved, with the root-mean-squared error of the fits decreasing from 3.0 mJy to 1.3 mJy.
Linear fits of comparison show that:\\ \\
$
S_{\rm Miller+2009} [{\rm mJy}] = (1.07\pm0.11) \times S_{\rm Ours} [{\rm mJy}]  
$\\ \\
$
S_{\rm FIRST} [{\rm mJy}] = (0.94\pm0.23) \times S_{\rm Ours} [{\rm mJy}]  
$\\ \\
$
S_{\rm Miller+2009} [{\rm mJy}] = (0.95\pm0.09) \times S_{\rm PyBDSF\ Miller} [{\rm mJy}]  
$\\ \\
Power-law fits confirm the linear relations between our flux density and the \citet[][]{Miller2009AJ....137.4436M} flux density (an index of $1.00\pm0.07$), and the FIRST flux density (an index of $1.00\pm0.06$).

\section{1.4 GHz continuum source counts}

\begin{figure}
    \includegraphics[width=1.00\columnwidth]{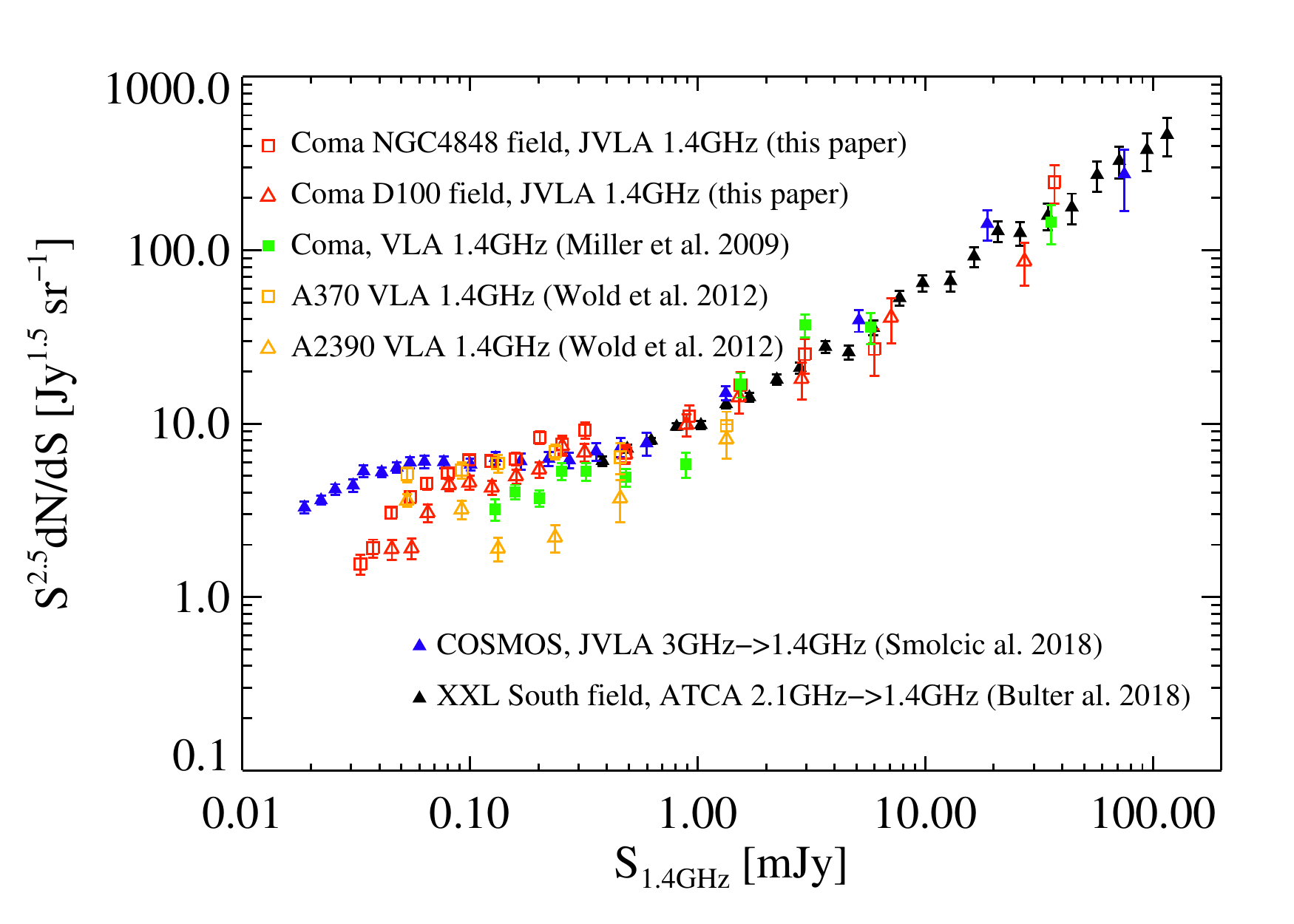}
    \vspace{-0.5cm}
    \caption{Comparison of source counts from our work and previous results.
For comparison, 3 GHz and 2.1 GHz flux densities from \citet{Smolcic2017A&A...602A...1S} and \citet{Butler2018A&A...620A..16B} are converted to 1.4 GHz flux densities assuming a spectral index of -0.7 and -0.75, respectively.
}
    \label{fig:counts-Sv}
\end{figure}

Our new data generated the largest 1.4 GHz source catalog in the Coma cluster. We compare source counts in our regions with those from previous work in Fig.~\ref{fig:counts-Sv}. The source count is defined as $S^{2.5}dN/dS/A$. $S$ is the source flux density, $N$ is the source number and $A$ is the observed sky area in steradians.
Our source count is restricted to the 1975 and 1173 sources identified by PyBDSF in two fields. 
The sky area, $A$, is different for each bin of flux density, $S$. For flux density $S$, we used the sky area for which the rms was less than $1/4\ S$, because only sources with a peak flux density greater than 4$\sigma$ are identified by PyBDSF.
In both fields, the rms increases with the distance from the center because of the primary beam correction. So the region which is sensitive enough to detect faint sources is smaller than the region to detect bright sources.
The uncertainties in source counts were derived from Poisson statistics.
Our results are consistent with other studies between 0.1 mJy and 110 mJy. For lower flux density, the source counts are lower than \citet{Smolcic2017A&A...602A...1S}. However, we need to keep in mind that 
the completeness and bias corrections used by \citet{Smolcic2017A&A...602A...1S} have not been applied to our data.
The correction could be as large as a factor of 3 for a source with a flux density of 5$\sigma$ \citep[Table 4 and Fig. 18 of][]{Smolcic2017A&A...602A...1S}. At the same time, some inconsistency could be caused by different software packages (e.g. PyBDSF in our work, BLOBCAT in \citet{Smolcic2017A&A...602A...1S,Butler2018A&A...620A..16B} and SAD in \citet{Miller2009AJ....137.4436M,Wold2012ApJS..202....2W}) and thresholds (source peaks greater than $S/N>4$ or $S/N>5$) used to identify sources in these studies. 

\section{Radio emission from Ultra-Diffuse Galaxies in the Coma cluster}

We also examined the radio emission from ultra-diffuse galaxies (UDGs) in the Coma cluster.
With the radio sources detected in \citet{Miller2009AJ....137.4436M}, \citet{Struble2018MNRAS.473.4686S} found no evidence for radio emission from UDGs.
We match the 3084 radio sources detected in our deep observations and the UDGs in the Coma cluster \citep{Koda2015ApJ...807L...2K,Yagi2016ApJS..225...11Y}.
There are 449 UDGs within our fields (field 1 + 2, out to the 10\% primary beam response level) that cover 4152.9 arcmin$^2$. 
We find that there is 1 match within a 2$\arcsec$ radius, 6 matches within a 5$\arcsec$ radius and 26 matches within a 10$\arcsec$ radius.
Following \citet{Struble2018MNRAS.473.4686S}, the expected number of random matches within an offset of $r$ is $\sim$1.2 for $r<$ 2$\arcsec$, $\sim$7.3 for $r<$ 5$\arcsec$, $\sim$29.0 for $r<$ 10$\arcsec$.
Thus, even with our much deeper radio data, there is no evidence of radio emission from UDGs in the Coma cluster.


\bsp	
\label{lastpage}
\end{document}